\documentclass[aps,pre,showpacs,twocolumn,superscriptaddress,
              floats,floatfix]{revtex4}
\usepackage{amssymb,amsmath}
\usepackage{graphics}
\usepackage{color}
\usepackage{subfigure}
\usepackage{epsfig}
\usepackage{bm}
\usepackage{ulem}

\def \epinj {\epsilon_{\rm inj}}
\def \epP {\epsilon_{\rm poly}}
\def \eppoly {\epsilon^{\rm p}_{\nu}}
\def \epflu {\epsilon^{\rm f}_{\nu}}

\def \rp {r_{\rm P}}
\def \bfu {{\bf u}}
\def \bomega {{\bm \omega}}

\def \Teddy {T_{\rm eddy}}

\def \taup  {\tau_{\rm P}}

\def \urms  {u_{\rm rms}}

\begin{document}
\title{Direct numerical simulations of statistically steady, homogeneous, 
isotropic fluid turbulence with polymer additives}
\author{Prasad Perlekar}
\email{p.perlekar@tue.nl}
\affiliation{Department of Mathematics and Computer Science, 
Eindhoven University of Technology, P.O. Box 513, 5600 MB Eindhoven, 
The Netherlands}
\author{Dhrubaditya Mitra}
\email{dhruba.mitra@gmail.com}
\affiliation{NORDITA, Roslagstullsbacken 23, 106 91 Stockholm,
Sweden}
\author{Rahul Pandit}
\email{rahul@physics.iisc.ernet.in}
\altaffiliation[\\ also at~]{Jawaharlal Nehru Centre For Advanced
Scientific Research, Jakkur, Bangalore, India}
\affiliation{Centre for Condensed Matter Theory,
Department of Physics,
Indian Institute of Science, Bangalore 560012, India.}
\begin{abstract}
We carry out a direct numerical simulation (DNS) study that reveals the 
effects of polymers on 
statistically steady, forced, homogeneous, isotropic fluid turbulence.  
We find clear manifestations of dissipation-reduction phenomena: 
On the addition of polymers to the turbulent fluid, we obtain a 
reduction in the energy dissipation rate, a significant modification 
of the fluid energy spectrum, especially in the deep-dissipation range, 
signatures of the suppression of small-scale structures, including a decrease 
in small-scale vorticity filaments. We also compare our results with
recent experiments and earlier DNS studies of decaying fluid turbulence
with polymer additives. 
\end{abstract}
\keywords{Turbulence, polymers}
\pacs{}
\maketitle
\section{Introduction}
The addition of small amounts of polymers to a turbulent fluid
leads to dramatic changes that include modifications of the
small-scale properties of the flow~\cite{hoy77,kal_poly04,per06}
and, in wall-bounded flows, the phenomenon of drag
reduction~\cite{toms49,lum73,vir75}, in which polymer additives
allow the maintenance of a given flow rate at a lower pressure
gradient than is required without these additives. Several
experimental, numerical, and analytical studies have investigated
drag reduction~\cite{vir75,lum73,too97,dub01,pta01,pta03,lvo04,ang05,pro08}
in wall-bounded flows. These studies have shown that the addition
of polymers modifies the turbulence significantly in the region
near the wall; and this leads to an increase in the mean velocity
in the bulk.  By contrast, there have been only some
investigations of the effects of polymer additives on
homogeneous, isotropic turbulence.  Examples include recent
experiments~\cite{oue09,lib05,lib06}, which have been designed to
obtain a high degree of isotropy in the turbulent flow, and
shell-model studies and direct numerical simulations 
(DNS)~\cite{ben03,vai03,kal_poly04,ang05,per06}.
These studies have shown that the addition of polymers to a turbulent
flow leads to a considerable reduction in small-scale structures;
and they have also discovered the phenomenon of dissipation
reduction, namely, a reduction in the energy dissipation rate
$\epsilon$, in decaying turbulence~\cite{kal_poly04,per06}.
In this paper we first elucidate the phenomenon of dissipation
reduction for the case of statistically steady, homogeneous, 
isotropic turbulence with polymer additives; we then study the 
small-scale properties of such flows.

We do this by conducting a series of high-resolution DNS studies
of the three-dimensional Navier-Stokes equation coupled to an equation for
the polymer conformation tensor, which describes the polymer additives at the
level of the FENE-P~\cite{kal_poly04,per06} model. Before we give the
details of our study, it is useful to summarise our principal results; these
are in two parts. 

The first part contains results from our DNS of forced,
statistically steady, fluid turbulence, with polymer additives,
at moderate Reynolds numbers ($Re_{\lambda}\simeq80$).  The
forcing is chosen such that the energy injected into the fluid
remains fixed~\cite{lam05}, both with and without polymers; this
mimics the forcing scheme used in the experiments of
Ref.~\cite{lib05,lib06}.  We find that, on the addition of
polymers, the energy in the statistically steady state and 
the energy-dissipation rate
are reduced. This dissipation reduction increases with an
increase in the polymer concentration $c$ at fixed Weissenberg
number $We$, the ratio (see Table~\ref{tablech3:para}) of the
polymer time scale $\taup$ to a shearing time scale in the
turbulent fluid. 
The dissipation reduction also increases with $We$ if we hold $c$
fixed.
The dissipation reduction seen in our simulations should
not be confused with the phenomenon of drag reduction seen
in wall-bounded flows. 
In the fluid energy spectrum we find that the 
energy content increases marginally at small wavevectors 
on the addition of polymers, but it decreases for intermediate wavevectors.
In this part of our study we use $256^3$ collocation points and
attain a moderate Reynolds  numbers $Re_{\lambda}\simeq80$; but we
do not resolve the deep-dissipation range.
(We consider the deep dissipation range 
in the following
paragraph.) We also obtain the structural properties of the fluid with and
without polymers and show that polymers suppress large-vorticity and
large-strain events; our results here are in qualitative agreement with the
experiments of Ref.~\cite{lib05,lib06}. Furthermore, we
find, as in our study of decaying turbulence~\cite{per06}, that
the polymer extension increases with an increase in the polymer
relaxation time $\taup$.
We compare our results, e.g., those for the energy spectrum, 
with their counterparts in our earlier study of decaying fluid turbulence 
with polymer additives~\cite{per06}. In such comparisons,
we use averages over the statistically steady state of our system here; 
and, for the case of decaying turbulence, we use data obtained 
at the cascade-completion time, at which a plot of the energy 
dissipation rate versus time displays a maximum.

In the second part of our study we carry out the
highest-resolution DNS, attempted so far, of forced,
statistically steady, fluid turbulence with polymer additives; we
drive the fluid by an external, stochastic force
as in Ref.~\cite{esw88}. This part of our study has been
designed to uncover the effects of polymers on the deep-dissipation range, 
so the Reynolds numbers is small ($Re_{\lambda}\simeq16$). By
comparing fluid energy spectra, with and without polymers, we
find that the polymers suppress the energy in the dissipation
range but increase it in the deep-dissipation range.  Finally, we
calculate the second-order velocity structure function $S_2(r)$
directly from the energy spectrum via a Fourier transformation;
this shows that $S_2(r)$ with polymers is smaller than $S_2(r)$
without polymers in this range.

The remaining part of this paper is organised as follows. In
Section~\ref{sec:mode} we present the equations we use for the polymer
solution and describe, in subsection~\ref{sec:nsch}, the method we use for
the numerical integration of these equations. Section~\ref{results} is
devoted to a discussion of our results; as we have mentioned above, these are
divided into two parts; the first part is contained in
subsections~\ref{sub:eeps}-\ref{sub:ss1} and the second in
subsection~\ref{sub:ekN512}. Section~\ref{conclusions} contains a concluding
discussion.

\section{Equations and Numerical Methods} 
\label{sec:mode}
We model a polymeric fluid solution by using the three-dimensional,
Navier-Stokes (NS) equations for the fluid coupled with the Finitely
Extensible Nonlinear Elastic-Peterlin (FENE-P) equation for the polymer
additives~\cite{per06}. The polymer contribution to the fluid is modelled by
an extra stress term in the NS equations. The FENE-P equation approximates a
polymer molecule by a nonlinear dumbbell, which has a single relaxation time
and an upper bound on the maximum extension.  The NS and FENE-P (henceforth
NSP) equations are 
\begin{eqnarray}
D_t{\bf u} &=& \nu \nabla^2 {\bf u}+
              \frac{\mu}{\taup}\nabla.[f(\rp){\cal C}] - {\nabla}p + {\bf f};  
                                                 \label{ch3ns}\\
D_t{\cal C}&=& {\cal C}. (\nabla {\bf u}) + 
                {(\nabla {\bf u})^T}.{\cal C} - 
                \frac{{f(\rp){\cal C} }- {\cal I}}{\taup}.
                                                   \label{ch3FENE}
\end{eqnarray}
Here ${{\bf u}({\bf x},t)}$ is the fluid velocity at point 
${\bf x}$ and time $t$, incompressibility is enforced by 
$\nabla .{\bf u}=0$, $D_t=\partial_t + {\bf u}.\nabla$, 
$\nu$ is the kinematic viscosity of the fluid, $\mu$ the viscosity 
parameter for the solute (FENE-P), $\taup$ the polymer relaxation time, 
$\rho$ the solvent density (set to $1$), $p$  
the pressure, ${{\bf f}({\bf x},t})$ the external force at point 
${\bf x}$ and time $t$, $(\nabla {\bf u})^T$ the transpose of 
$({\nabla {\bf u}})$, ${\cal C}_{\alpha\beta}\equiv 
             {\langle{R_\alpha}{R_\beta}\rangle}$ the 
elements of the polymer-conformation tensor ${\cal C}$ (angular 
brackets indicate an average over polymer configurations), 
${\cal I}$ the identity tensor with elements $\delta_{\alpha \beta}$, 
$f(\rp)\equiv{(L^2 -3)/(L^2 - \rp^2)}$ the FENE-P potential that ensures 
finite extensibility, $\rp \equiv \sqrt{Tr(\cal C)}$ and $L$ the
length and the maximum possible extension, respectively, of the polymers, 
and $c\equiv\mu/(\nu+\mu)$ a dimensionless measure of the polymer 
concentration \cite{vai03}; $c=0.1$ corresponds, roughly, 
to $100$ppm for polyethylene oxide \cite{vir75}. Table~\ref{tablech3:para} 
lists the parameters of our simulations.
\begin{table}[!h]
\begin{center}
   \begin{tabular}{@{\extracolsep{\fill}} c c c c c c c c}
    \hline
    $ $ &$N$ & $\delta{t}$ & $L$ & $\nu$ & $\taup$ & $c$ & $We$\\
   \hline \hline
    {\tt NSP-256A} & $256$ &  $5.0\times10^{-4}$ & $100$ &  $5\times10^{-3}$  &    $0.5$  & $~0.1$ & $3.5$ \\
    {\tt NSP-256B} & $256$ &  $5.0\times10^{-4}$ & $100$ &  $5\times10^{-3}$  &    $1.0$  & $~0.1$ & $7.1$\\
    {\tt NSP-512}  & $512$  &  $10^{-3}$ & $100$ &  $5\times10^{-2}$  & $1.0$ & $0.1$ & $0.9$ \\
\hline
\end{tabular}
\end{center}
\caption{\small The cube root $N$ of the number of collocation points, the
time step $\delta t$, the maximum possible polymer extension $L$, the
kinematic viscosity $\nu$, the polymer-relaxation time $\taup$, and the
polymer concentration parameter $c$ for our four runs ${\tt NSP-256A}$, ${\tt
NSP-256B}$ and ${\tt NSP-512}$.  We also carry out DNS studies of the NS
equation with the same numerical resolutions as in our NSP runs. The
Taylor-microscale Reynolds number $Re_{\lambda}\equiv\sqrt{20}{\mathcal
E^{\rm f}}/\sqrt{3\nu\epflu }$ and the Weissenberg number $We \equiv
\taup\sqrt{\epflu/\nu}$ are as follows: ${\tt NSP-256A}$ and ${\tt
NSP-256B}$: $Re_{\lambda} \simeq 80$ and  ${\tt NSP-512}$:
$Re_{\lambda}\simeq16$; the Kolmogorov dissipation length scale
$\eta\equiv(\nu^3/\epflu)^{1/4}$. For our runs ${\tt NSP-256A-B}$,
$\eta\simeq1.07\delta x$; and for run ${\tt NSP-512}$,  $\eta \simeq
19\delta x$,  where $\delta x \equiv {\mathbb L}/N$ is the grid resolution of
our simulations. The integral length scale $l_{\rm int}\equiv(3\pi/4)\sum
k^{-1} E(k)/(\sum E(k))$ and $\Teddy\equiv \urms/l_{\rm int}$ are as follows:
${\tt NSP-256A}$ and ${\tt NSP-256B}$: $l_{\rm int} \simeq 1.3$ and $\Teddy
\simeq 1.2$; and, for ${\tt NSP-512}$, $l_{\rm int}\simeq 2.05$ and $\Teddy
\simeq 4.0$.}
\label{tablech3:para}
\end{table} 
\subsection{Numerical Methods}
\label{sec:nsch}
We consider homogeneous, isotropic, turbulence, so we use
periodic boundary conditions and solve Eq.~(\ref{ch3ns}) by using
a pseudospectral method~\cite{Can88,vin91}. We use $N^3$
collocation points in a cubic domain (side ${\mathbb L}=2\pi$).
We eliminate aliasing errors by the 2/3 rule~\cite{Can88,vin91},
to obtain reliable data at small length scales; and we use a
second-order, slaved, Adams-Bashforth scheme for time marching.
In earlier numerical studies of homogeneous, isotropic turbulence
with polymer additives it has been shown that sharp gradients are
formed during the time evolution of the polymer conformation tensor;
this can lead to dispersion errors~\cite{vai03,vai06}. To avoid
these dispersion errors, shock-capturing schemes have been used
to evaluate the polymer-advection term $[{(\bfu \cdot \nabla)
{\mathcal C}}]$ in Ref.~\cite{vai06}.  In our simulations we have
modified the Cholesky-decomposition scheme of Ref.~\cite{vai03},
which preserves the symmetric positive definite nature of the
tensor ${\mathcal C}$. We incorporate the large gradients of the
polymer conformation tensor by
evaluating the polymer-advection term $[{(\bfu \cdot \nabla)
\ell}]$ via the Kurganov-Tadmor shock-capturing scheme \cite{kt00}.  For the
derivatives on the right-hand side of Eq.~(\ref{ch3FENE}) we use
an explicit, fourth-order, central-finite-difference scheme in
space; and the temporal evolution is carried out by using an
Adams-Bashforth scheme.  The numerical error in $\rp$ must be
controlled by choosing a small time step $\delta t$, otherwise
$\rp$ can become larger than $L$, which leads to a numerical
instability; this time step is much smaller than what is
necessary for a pseudospectral DNS of the NS equation alone.
Table~\ref{tablech3:para} lists the parameters we use. We
preserve the symmetric-positive-definite (SPD) nature of $\cal C$
at all times by using~\cite{vai03} the following
Cholesky-decomposition scheme: If we define
\begin{equation}
{\cal J} \equiv f(\rp) {\cal C}, 
\end{equation}
~Eq.~(\ref{ch3FENE}) becomes 
\begin{equation} 	
D_t{\cal J} = {\cal J}. (\nabla {\bf u}) 
+ ({\nabla \bf u})^T .{\cal J} -s({\cal J} - {\cal I})+ q {\cal J},
\label{conj} 
\end{equation} 
where 
\begin{eqnarray}
\nonumber
s&=&\frac{L^2 -3+ j^2}{\taup L^2}, \\
\nonumber
q&=&\frac{d/(L^2 -3)-(L^2 -3+ j^2)(j^2 -3)}{(\taup L^2(L^2 -3))}, \\
\nonumber
j^2&\equiv& Tr({\cal J}), {\rm and} \\
\nonumber
d &=& Tr[ {\cal J}. (\nabla{\bf u}) + (\nabla{\bf u})^T .{\cal J}].
\end{eqnarray}
${\cal C}$ and hence ${\cal J}$ are SPD matrices; we can, therefore, write
${\cal J}= {\cal LL}^T$, where ${\cal L}$ is a lower-triangular matrix
with elements $\ell_{ij}$, such that $\ell_{ij}=0$ for $j>i$, and
\begin{equation}
 \mathcal{J} \equiv \left( \begin{array}{ccc}
 \ell_{11}^2 & \ell_{11}\ell_{21} & \ell_{11}\ell_{31}\\
 \ell_{11}\ell_{21} & \ell_{21}^2+\ell_{22}^2 & \ell_{21}\ell_{31}
 +\ell_{22}\ell_{32}\\
 \ell_{11}\ell_{31} & \ell_{21}\ell_{31}+\ell_{22}\ell_{32}
 & \ell_{31}^2+\ell_{32}^2+\ell_{33}^2
 \end{array} \right) .
 \end{equation}
Equation~\eqref{conj} now yields $(1\le i \le3$ and
$\Gamma_{ij}=\partial_i u_j )$ the following set of equations:
\begin{eqnarray}
\nonumber
{D_t \ell_{i1}}  &=& \sum_k \Gamma_{ki}\ell_{k1}
+ \frac{1}{2}\Big[(q-s)\ell_{i1}+(-1)^{(i \bmod 1)}
\frac{s\ell_{i1}}{\ell^2_{11}}\Big] \\    
\nonumber    
&&{}+ (\delta_{i3}+\delta_{i2})\frac{\ell_{i2}}{\ell_{11}}
 \sum_{m>1}\Gamma_{m1}\ell_{m2} \\
\nonumber
&&{} + \delta_{i3}\Gamma_{i1}\frac{\ell^2_{33}}{\ell_{11}},~
\mbox{for}~i\geq1; \\
\nonumber
{ D_t \ell_{i2}}  &=& \sum_{m\geqslant2}\Gamma_{mi}\ell_{m2}
 -\frac{\ell_{i1}}{\ell_{11}}
  \sum_{m\geqslant2}\Gamma_{m1}\ell_{m2}\\
\nonumber
&&{} + \frac{1}{2}\Big[(q-s)\ell_{i2}+(-1)^{(i+2)}
 s\frac{\ell_{i2}}{\ell_{22}^2}
\Big(1+\frac{\ell^2_{21}}{\ell^2_{11}}\Big)\Big]\\
\nonumber          
&&{}+ \delta_{i3}\Big[\frac{\ell^2_{33}}{\ell_{22}}
\Big(\Gamma_{32}-\Gamma_{31}\frac{\ell_{21}}{\ell_{11}}\Big)
 + s\frac{\ell_{21}\ell_{31}}{\ell^2_{11}\ell_{22}}\Big],~
\mbox{for}~i\geq2;\\
\nonumber 
{ D_t \ell_{33}} &=& \Gamma_{33} \ell_{33}
   - \ell_{33}\Big[\sum_{m<3}\frac{\Gamma_{3m}\ell_{3m}}{\ell_{mm}}\Big] 
     + \frac{\Gamma_{31}\ell_{32}\ell_{21}\ell_{33}}{\ell_{11}\ell_{22}} \\
   \nonumber
    &&{}-s\frac{\ell_{21}\ell_{31}\ell_{32}}{\ell^2_{11}\ell_{22}\ell_{33}}
       + \frac{1}{2}\Big[(q-s)\ell_{33} \\
    &&{}   + \frac{s}{\ell_{33}}  
        \Big(1+\sum_{m<3}\frac{\ell^2_{3m}}{\ell^2_{mm}}\Big)
       +\frac{s\ell^2_{21}\ell^2_{32}}{\ell^2_{11}\ell^2_{22}\ell_{33}} \Big].
\label{ellij}
\end{eqnarray}
The SPD nature of  $\cal C$ is preserved by Eqs.~\eqref{ellij} if
$\ell_{ii} > 0$, which we enforce explicitly~\cite{vai03} by
considering the evolution of $\ln(\ell_{ii})$ instead of $\ell_{ii}$. 

We resolve the sharp gradients in the polymer conformation tensor  by
discretizing the polymer advection term by using the Kurganov-Tadmor 
scheme \cite{kt00}. Below we show the discretization of the advection 
term $u \partial_x \ell$, where ${\bf u}\equiv(u,v,w)$ and $\ell$ is one 
of the components of the $\ell_{\alpha \beta}$; the discretization of the 
other advection terms in Eq.~(\ref{ellij}) is similar.   
\begin{eqnarray}
u \partial_x \ell &=& \frac{H_{i+1/2,j,k}-H_{i-1/2,j,k}}{\delta x}, 
\nonumber \\
H_{i+1/2,j,k} &=& \frac{u_{i+1/2,j,k}[\ell_{i+1/2,j,k}^{+} + \ell_{i+1/2,j,k}^{-}]}{2} 
\nonumber \\
	&&-\frac{a_{i+1/2,j,k}[\ell_{i+1/2,j,k}^{+}-\ell_{i+1/2,j,k}^{-}]}{2}, 
\nonumber \\
\ell_{i+1/2,j,k}^{\pm} &=& \ell_{i+1,j,k}\mp \frac{\delta x}{2}(\partial_x \ell)_{i+1/2\pm1/2,j,k} \;, \nonumber \\
a_{i+1/2,j,k} &\equiv& |u_{i+1/2,j,k}|\;,
\end{eqnarray} 
where $i,j,k=0,\ldots (N-1)$ denote the grid points and $\delta x = \delta y = \delta z$ is the grid spacing  along the three directions.

We use the following initial conditions (superscript $0$): ${\cal
C}^0_{mn}({\bf x}) =  \delta_{mn}$  for all ${\bf x}$; and
${u}^0_m({\bf k})= P_{mn}({\bf k}){v}^0_n({\bf k})
\exp(\iota{\theta_n(\bf{k})})$, with $m,n=x,y,z$,
$P_{mn}=(\delta_{mn}-k_mk_n/k^2)$ the transverse projection
operator, ${\bf k}$ the wave vector with components $k_m=
(-N/2,-N/2+1,\ldots,N/2)$ and magnitude  $k=|{\bf k}|$,
$\theta_n({\bf k})$ random numbers distributed uniformly between
$0$ and $2\pi$, and $v^0_n({\bf k})$ chosen such that the initial
kinetic-energy spectrum is  $E^0(k) = k^4 \exp(-2.0 {k}^2)$. This
initial condition corresponds to a state in which the fluid
energy is concentrated, to begin with, at small $k$ (large length
scales); and the polymers are in a coiled state. Our simulations
are run for $45\Teddy$ and a statistically steady state is reached in roughly
$10\Teddy$, where the integral-scale, eddy-turnover time $\Teddy
\equiv \urms/l_{\rm int}$, with $\urms$ the root-mean-square
velocity and $l_{\rm int}\equiv\sum_{k}k^{-1}E(k)/\sum_{k}E(k)$ the
integral length scale. Along with our runs ${\tt NSP-256A}$ and ${\tt
NSP-256B}$ we also carry out pure-fluid, NS simulations till a statistically 
steady state
is reached; this takes about $10-15\Teddy$. Once this pure-fluid
simulation reaches a statistically steady state, we add polymers to the fluid 
at $27\Teddy$; i.e., beyond this time we solve the coupled NSP 
equations \ref{ch3ns} and \ref{ch3FENE} by using the methods given 
above. We then allow $5-6\Teddy$ to elapse, so that transients die down, and
then we collect data for fluid and polymer statistics for another
$25\Teddy$ for our runs ${\tt NSP-256A}$ and ${\tt NSP-256B}$. 
\section{Results}
\label{results}
We now present the results that we have obtained from our DNS.  In addition
to ${\bf u}({\bf x},t)$, its Fourier transform ${\bf u}_{\bf k}(t)$, and
${\cal C}({\bf x},t)$, we monitor the vorticity $\bomega \equiv \nabla \times
{\bf u}$, the kinetic-energy spectrum $E(k,t)\equiv\sum_{k-1/2< k' \le
k+1/2}|{\bf u}^2_{\bf k'}(t)|$, the total kinetic energy ~${\mathcal E}(t)
\equiv\sum_kE(k,t)$, the energy-dissipation-rate $\epsilon_{\nu}(t) \equiv \nu
\sum_k k^2 E(k,t)$, the probability distribution of scaled polymer extensions
$P(\rp^2/L^2)$, the PDF of the strain and the modulus of the vorticity, and
the eigenvalues of the strain tensor. For notational convenience, we do not
display the dependence on $c$ explicitly.  In subsection~\ref{sub:eeps} we
present the time evolution of $E$ and $\epsilon_{\nu}$ and provide evidence for
dissipation reduction by polymer additives. This is followed by
subsections~\ref{sub:esp1} and \ref{sub:ss1} that deal, respectively, with
the effects of polymers on fluid energy spectra and small-scale structures in
turbulent flows. In subsection~\ref{sub:ekN512} we examine the modification,
by polymer additives, of fluid-energy spectra in the deep dissipation range.
\subsection{The Energy and its Dissipation Rate}
\label{sub:eeps}
\begin{figure}[!h]
\begin{center}
\includegraphics[width=\linewidth]{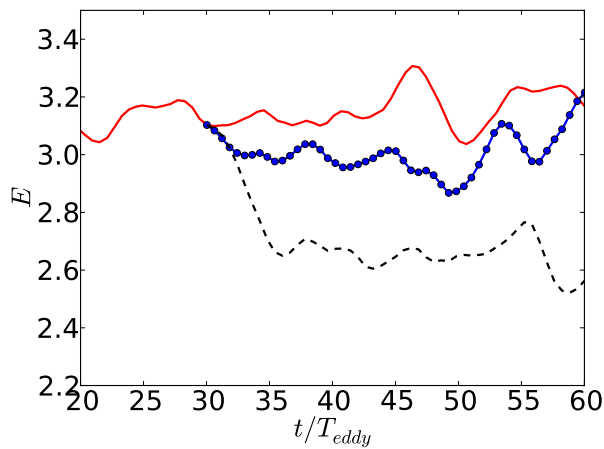}
\caption{\label{figch3:enf}(Color online) A plot of the 
fluid energy $E$ versus the dimensionless time $t/\Teddy$ 
(runs {\tt NSP-256A} and {\tt NSP-256B}) for 
Weissenberg numbers $We = 3.5$ ({\textcolor{blue}{blue circles}}) and 
$We = 7.1$ (black dashed line). The corresponding plot for 
the pure-fluid case is also shown for comparison~(\textcolor{red}{red line}). 
The polymers are added to the fluid at $t=27\Teddy$.
} 
\end{center}
\end{figure}

We first consider the effects of polymer additives on the time evolution of
the fluid energy $E$ for our runs ${\tt NSP-256A}$ and ${\tt NSP-256B}$; this
is shown in Fig.~\ref{figch3:enf}. The polymers are added to the fluid at
$t=27\Teddy$.  The addition of polymers leads to a new statistically steady
state; specifically, we find for $We=3.5$ and $We=7.1$, that the average
energy of the fluid with polymers is reduced in comparison to the average
energy of the fluid without polymers. 
By using Eq.~\eqref{ch3ns}, we obtain the following
energy-balance equation for the fluid with polymer additives:
\begin{eqnarray}
\frac{dE}{dt} &=& \epsilon_\nu + \epP + \epinj, \\
\nonumber \\
\epsilon_\nu &=& -\nu\frac{1}{V}\int {\bf u} \cdot \nabla^2 {\bf u}, \nonumber\\
\epP  &=&  (\frac{\mu}{\taup})\left\{\frac{1}{V} \int {\bf u} \cdot 
						\nabla [f(\rp){\mathcal C}]\right\},
\nonumber  \\
\epinj  &=& \frac{1}{V} \int {\bf f}\cdot{\bf u}\;.
\end{eqnarray}
In the statistically steady state $\frac{dE}{dt}=0$ and the energy injected 
is balanced by the fluid dissipation rate $\epsilon_\nu$ and the polymer 
dissipation $\epP$. Our simulations are designed to keep the energy injection
fixed. Therefore, we can determine how the dissipation gets
distributed between the fluid and polymer subsystems in forced, statistically
steady turbulence. 

\begin{figure}[!h]
\begin{center}
\includegraphics[width=\linewidth]{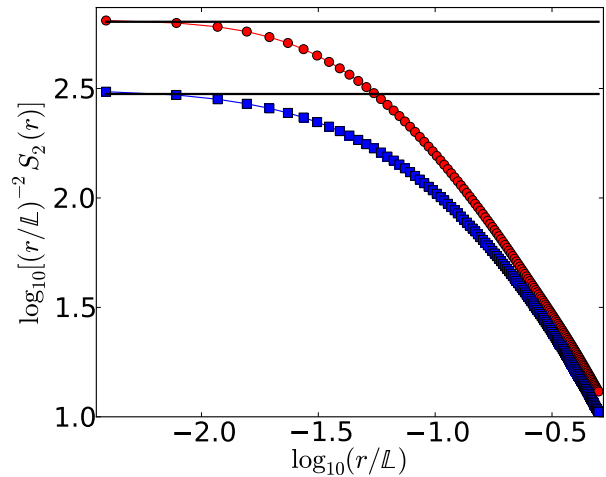}
\caption{\label{figch3:s2r}(Color online) Log-log (base 10)
plots of the second-order structure function $S_2(r)$, 
compensated by $(r/\mathbb{L})^{-2}$, versus $r/\mathbb{L}$, for our run 
{\tt NSP-256B} (\textcolor{blue}{blue square}) and  
for the pure-fluid case (\textcolor{red}{red circle}). The 
regions in which the horizonal black lines overlap with the points 
indicate the $r^2$ scaling ranges. 
} 
\end{center}
\end{figure}

\begin{figure}[!h]
\begin{center}
\includegraphics[width=\linewidth]{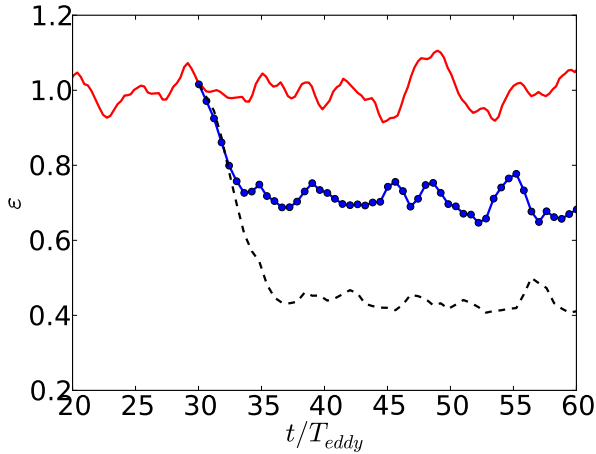}
\caption{\label{figch3:ediss}{ (Color online) Plots of the energy dissipation 
rate $\epsilon_\nu$ versus $t/\Teddy$ (runs {\tt NSP-256A} and 
{\tt NSP-256B}) for Weissenberg numbers $We = 3.5$ 
({\textcolor{blue}{blue circles}}) and 
$We = 7.1 ($black dashed line$)$. The corresponding plot for the pure 
fluid case is also shown for comparison~(\textcolor{red}{red line}). 
The polymers are added to the fluid at $t=27\Teddy$.}
} 
\end{center}
\end{figure}

\begin{figure}[!h]
\includegraphics[width=\linewidth]{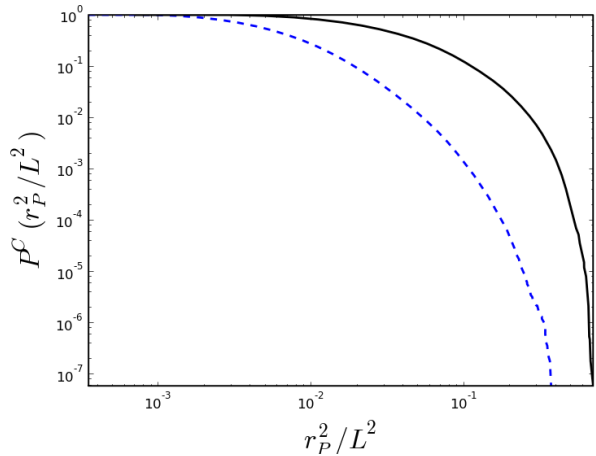}
\caption{\label{figch3:polyext}{(Color online) Log-log (base 10)
plots of the cumulative PDF $P^C(\rp^2/L^2)$ versus the scaled polymer 
extension $\rp^2/L^2$ for $We=3.5$ 
(blue dashed line for run {\tt NSP-256A}) and 
$We=7.1$ (full black line for run {\tt NSP-256B}). Note that
as $We$ increases so does the extension of the polymers. 
These plots are obtained from polymer configurations at $t=60\Teddy$.}} 
\end{figure}

Before we present our results for the kinetic-energy dissipation rate,
we first calculate the second order-structure function $S_2(r)$ 
via the following exact relation~\cite{Bat53}, 
\begin{equation}
S_2(r) = \int_{0}^{\infty} \left[1 - \frac{\sin(kr)}{kr} \right] E(k) dk. 
\label{eqch3:s2r}
\end{equation}
In Fig.~\ref{figch3:s2r} we give a log-log plot of  
$S_2(r)$, compensated by $(r/\mathbb{L})^{-2}$, as a function of 
$r/\mathbb{L}$. We find that, for small $r$, $S_2(r) \sim r^2$,
which implies that our DNS resolves the analytic range, which
follows from a Taylor expansion, of $S_2(r)$~\cite{sch+sre+yak89};
this guarantees that energy-dissipation rate has been calculated
accurately. 
In Fig.~\ref{figch3:ediss} we present plots of
$\epsilon_\nu(t)$ versus $t/\Teddy$ for $We=3.5$ and $We=7.1$ with the
polymer concentration $c=0.1$. We find that the average value of 
$\epsilon_\nu$ decreases as we increase $We$. 
This suggests the following natural definition of the percentage 
dissipation reduction for forced, homogeneous, isotropic turbulence:    
\begin{eqnarray}
\begin{aligned}
{\rm DR}\equiv\left(\frac{\langle\epflu\rangle-\langle\eppoly\rangle}
{\langle\epflu\rangle}\right)\times 100\%;
\end{aligned}
\label{ch3dragreduction}
\end{eqnarray}
here (and henceforth) the superscripts f and p stand, respectively, for
the fluid without and with polymers;
and the angular brackets
denote an average over the statistically steady state.
Percentage dissipation reduction, 
 ${\rm DR}$, rises with $We$;
this indicates that $\eppoly$ increases with
$We$. 
\footnote{The analog of dissipation reduction 
has been seen in a recent DNS of polymer-laden turbulence under uniform 
shear~\cite{rob+vai+col+bra10}.} 
Thus, in contrast to the trend we observed in our
decaying-turbulence DNS~\cite{per06}, ${\rm DR}$ increases 
with $We$: 
For $We=3.5$, ${\rm DR\simeq30\%}$ and,
for $We=7.1$, ${\rm DR\simeq50\%}$. 
Our interpretation is that this increase of ${\rm DR}$ with $We$ 
arises because the polymer extensions and, therefore, the
polymer stresses are much stronger in our forced-turbulence DNS than in 
our decaying-turbulence DNS 
(at least for the Reynolds numbers that we achieved in 
Ref.~\cite{per06}).
In Fig.~\ref{figch3:polyext} 
we show the cumulative PDF of the scaled polymer extension; this shows 
clearly that the extension of the polymers increases with $We$.
In general, the calculation of PDFs from numerical data is
plagued by errors originating from the binning of the data to make
histograms. Here instead we have used the rank-order method 
to calculate the corresponding cumulative PDF which
is free of binning errors~\cite{mit05a}.
\subsection{Energy spectra}
\label{sub:esp1}
\begin{figure}[!h]
\begin{center}
\includegraphics[width=\linewidth]{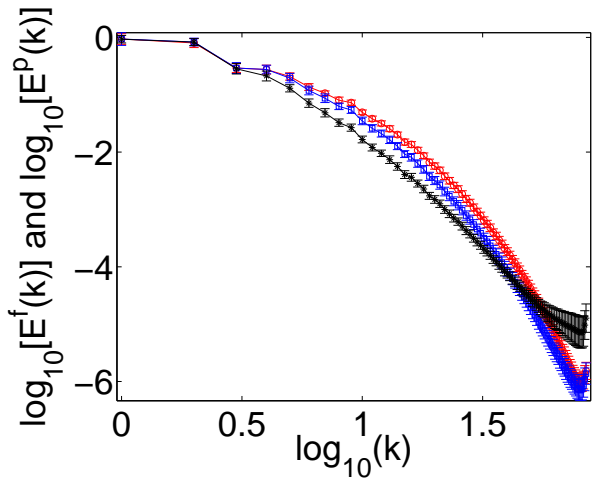}
\caption{\label{figch3:decspecb}{(Color online) Log-log plots (base 10) of the 
energy spectra $E^{\rm p}(k)$ versus 
$k$ (runs {\tt NSP-256A} and {\tt NSP-256B}) 
for $c=0.1$ and $We=3.5$~(blue squares) or 
$We=7.1$~(black stars); we give two-standard-deviation 
error bars. The corresponding pure-fluid spectrum  $E^{\rm f}(k)$ (red circles) 
is shown for comparison.}}
\end{center}
\end{figure}

In this subsection we study 
fluid-energy spectra $E^{\rm p}(k)$,
in the presence of polymer additives, for two different values of
the Weissenberg number $We$  and fixed  polymer concentration 
$c=0.1$ (Fig. \ref{figch3:decspecb}). We find that the energy content at
intermediate wave-vectors decreases with an increase in $We$. At
small wave-vector magnitudes $k$, we observe a small increase in
the spectrum on the addition of the polymers, but this increase
is within our numerical, two-standard-deviation error bars.  
Because of the moderate resolution of our simulations we are not able 
to resolve the dissipation range fully in these simulations. We address this 
issue by conducting high-resolution, low-Reynolds-number simulations in 
Sec.~\ref{sub:ekN512}.
\subsection{Small-scale structures}
\label{sub:ss1}
\begin{figure}[!h]
\begin{center}
\includegraphics[width=0.8\linewidth]{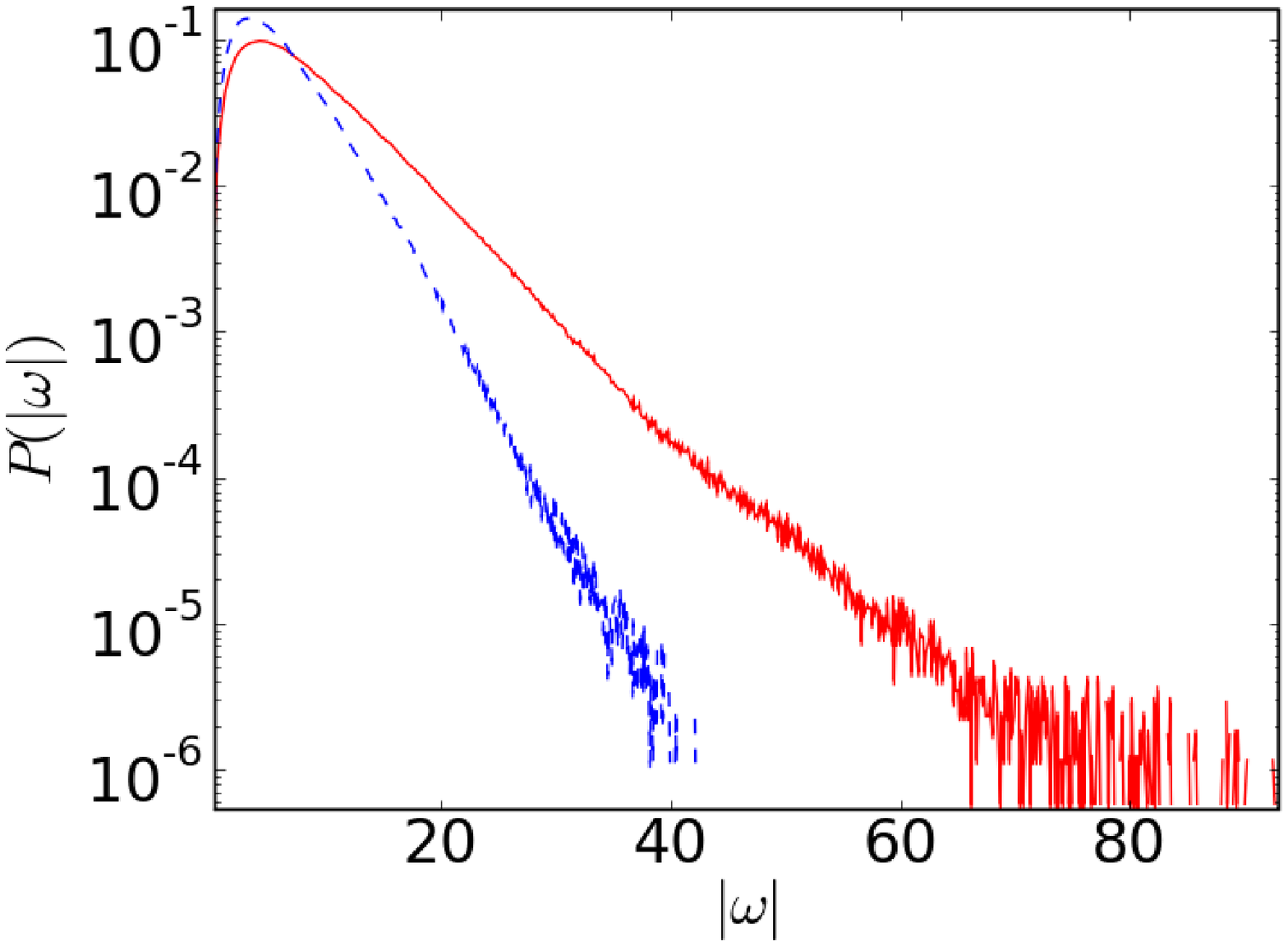}
\includegraphics[width=0.8\linewidth]{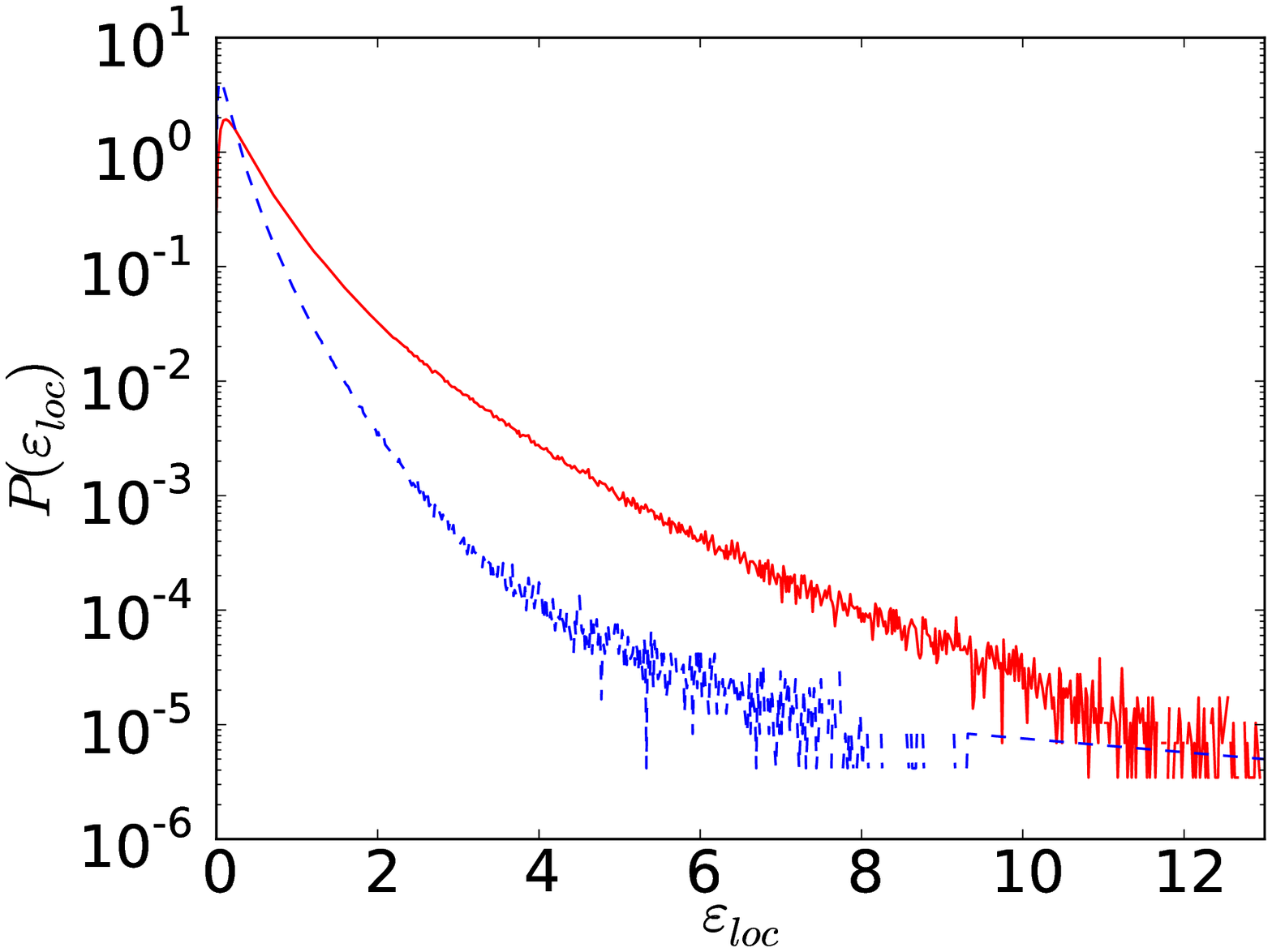}
\end{center}
\caption{\label{figch3:pdfve}{(Color online) Semilog plots (base 10) of the 
PDFs $P(|\bomega|)$ versus $|\bomega|$ (top panel) and 
$P(\epsilon_{\rm loc})$ versus $\epsilon_{\rm loc}$ (bottom panel), for our 
run {\tt NSP-256B}, with [$c=0.1$, $We=7.1$ (blue, dashed 
line)] and without [$c=0$ (full, red line)] polymer additives.}}
\end{figure}

\begin{figure}[!h]
\begin{center}
\includegraphics[width=0.8\linewidth]{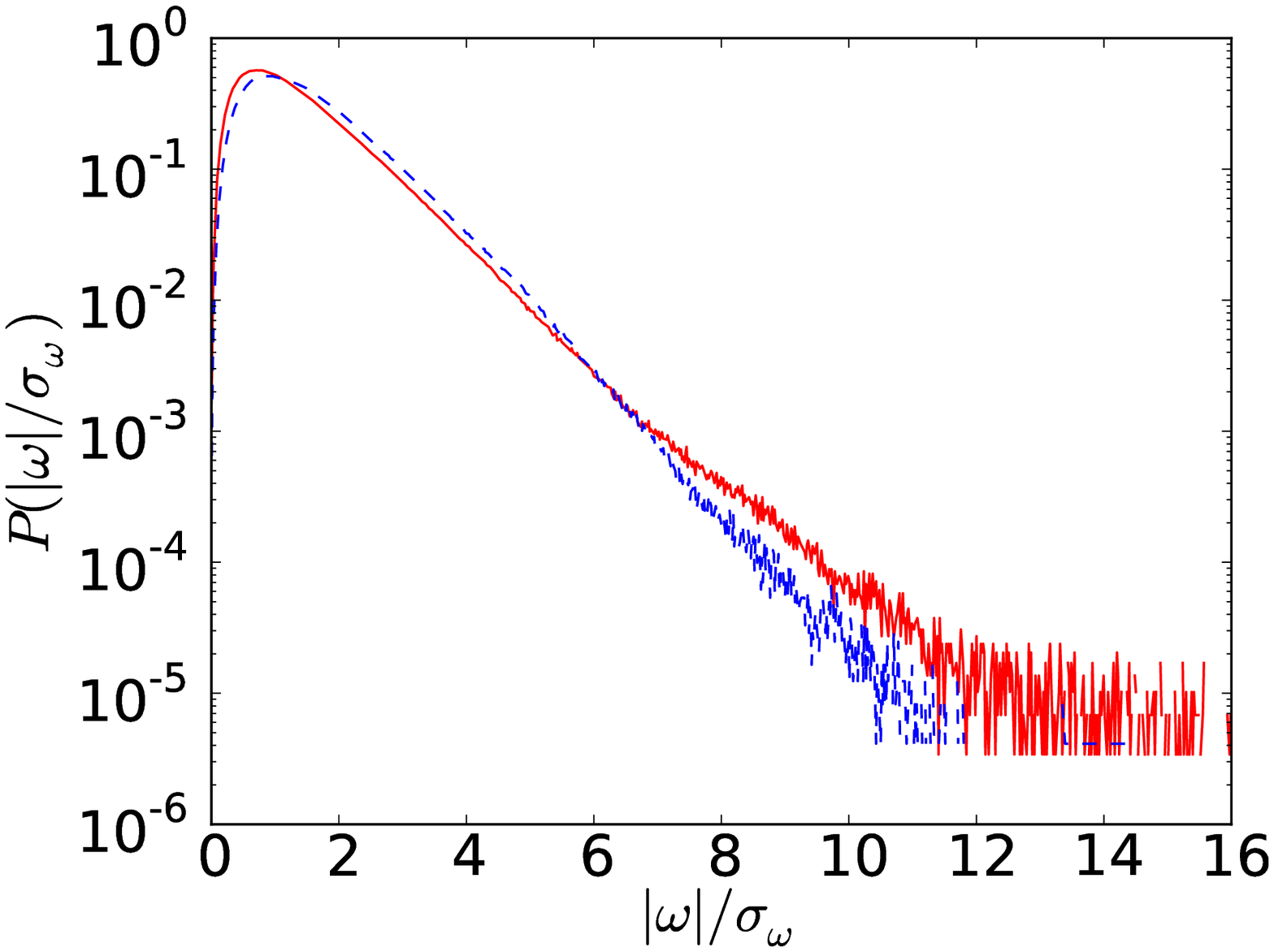}
\includegraphics[width=0.8\linewidth]{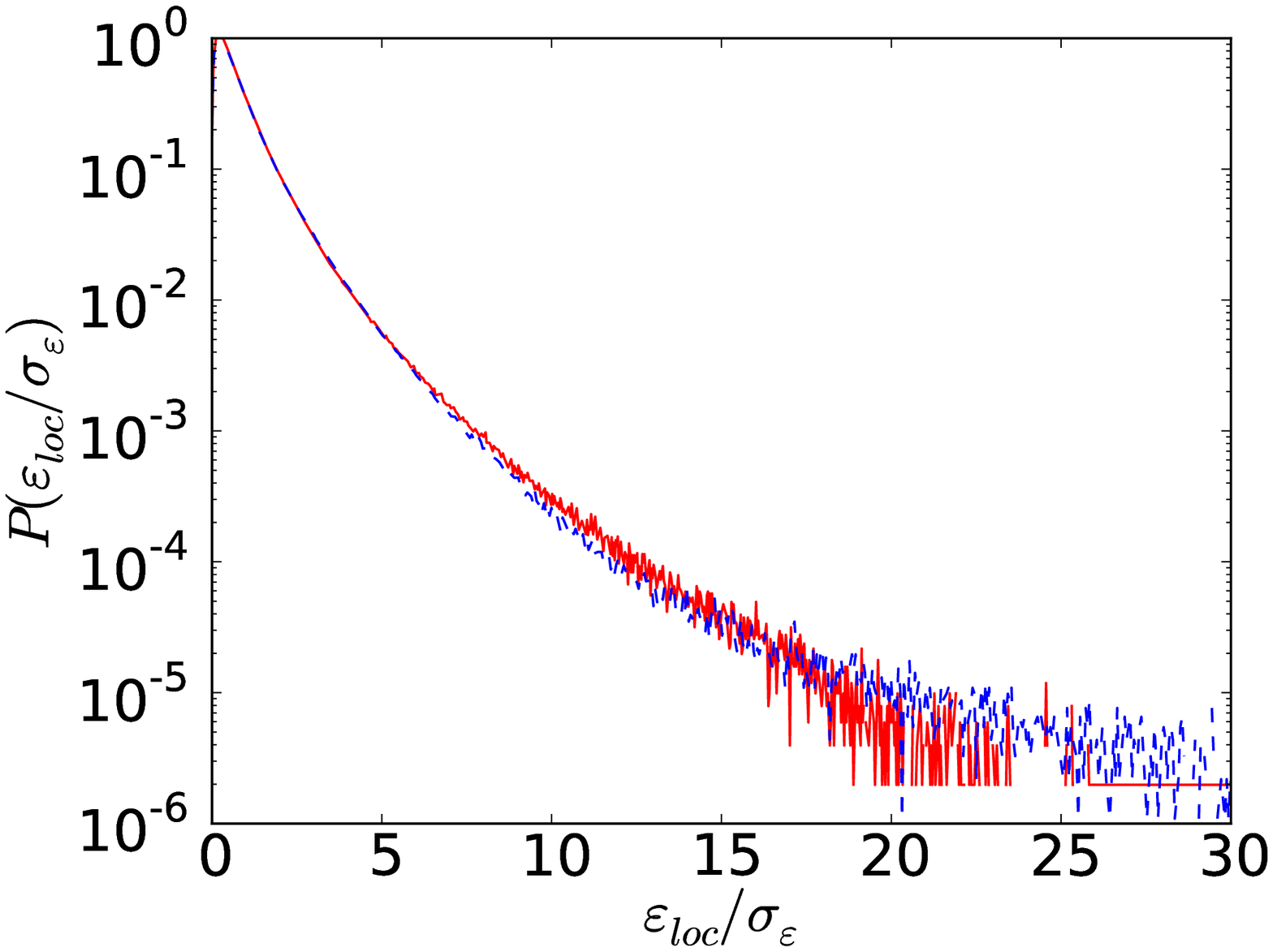}
\end{center}
\caption{\label{figch3:pdfnorm}{(Color online) Semilog plots (base 10) of the scaled PDFs 
$P(|\bomega|/\sigma)$ versus $|\bomega|/\sigma_\omega$ (top panel) and  
$P(\epsilon_{\rm loc}/\sigma_\epsilon)$ 
versus $\epsilon_{\rm loc}/\sigma_\epsilon$ (bottom panel), where $\sigma_\omega$ 
and $\sigma_\epsilon$ are the standard deviations for $|\bomega|$ and
$\epsilon_{\rm loc}$, respectively, for our run {\tt NSP-256B}, with [$c=0.1$, 
$We=7.1$ (\textcolor{blue}{dashed line})] and without 
[$c=0$ (\textcolor{red}{line})] polymer additives. These plots are normalized 
such that the area under each curve is unity.}}
\end{figure}

\begin{figure}[!h]
\begin{center}
\includegraphics[width=0.45\linewidth]{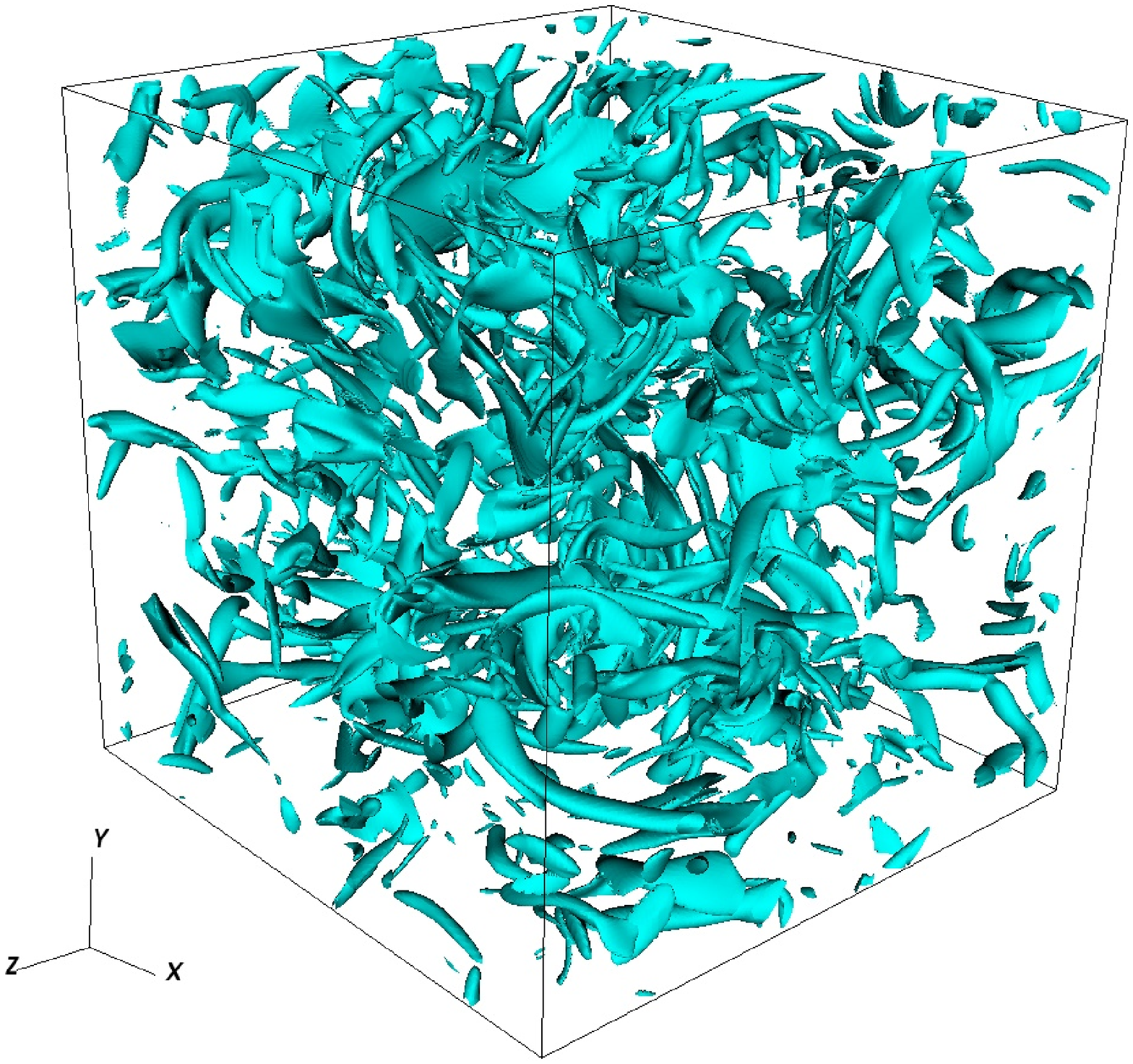}\\
\includegraphics[width=0.45\linewidth]{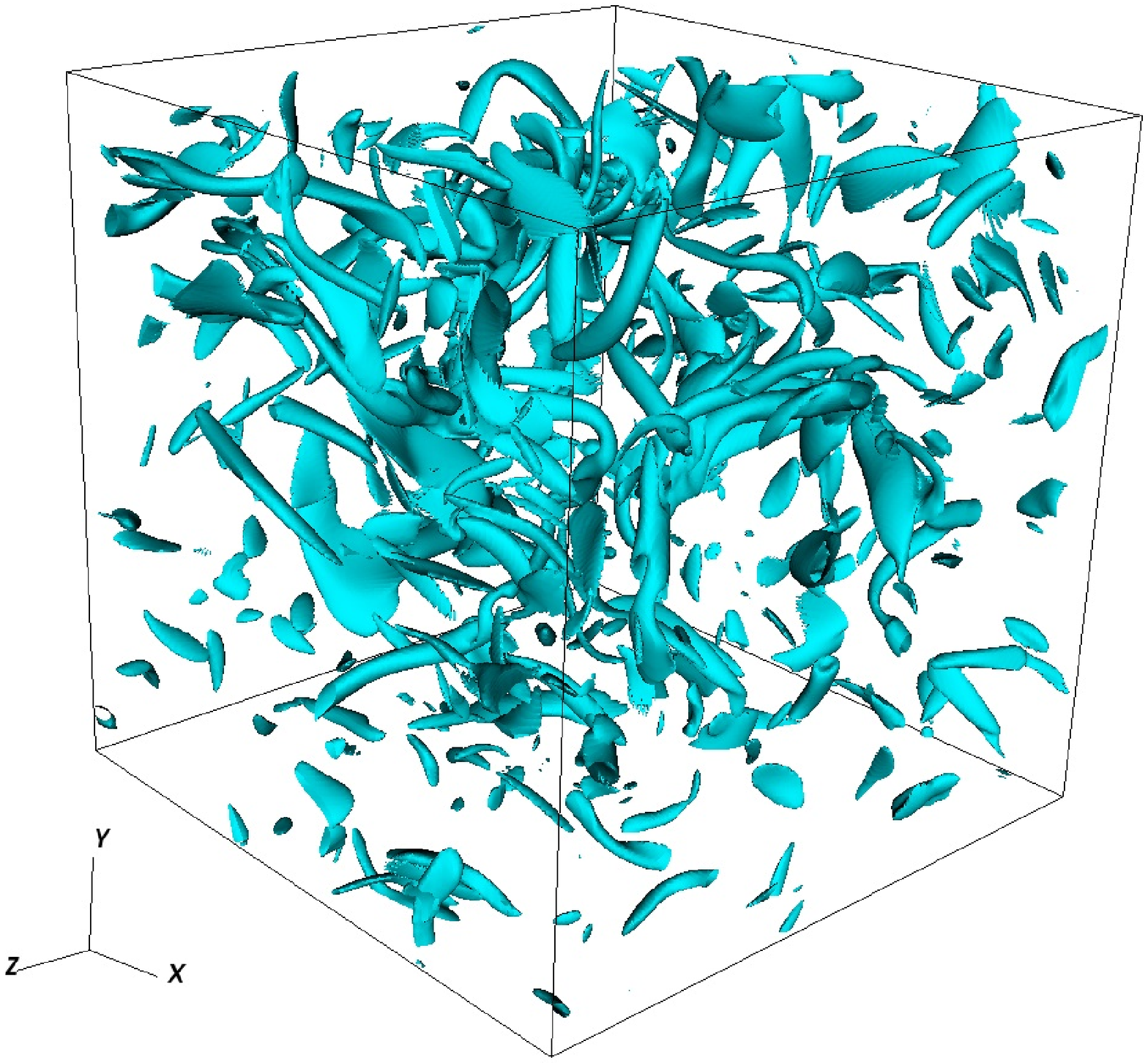}\\
\includegraphics[width=0.45\linewidth]{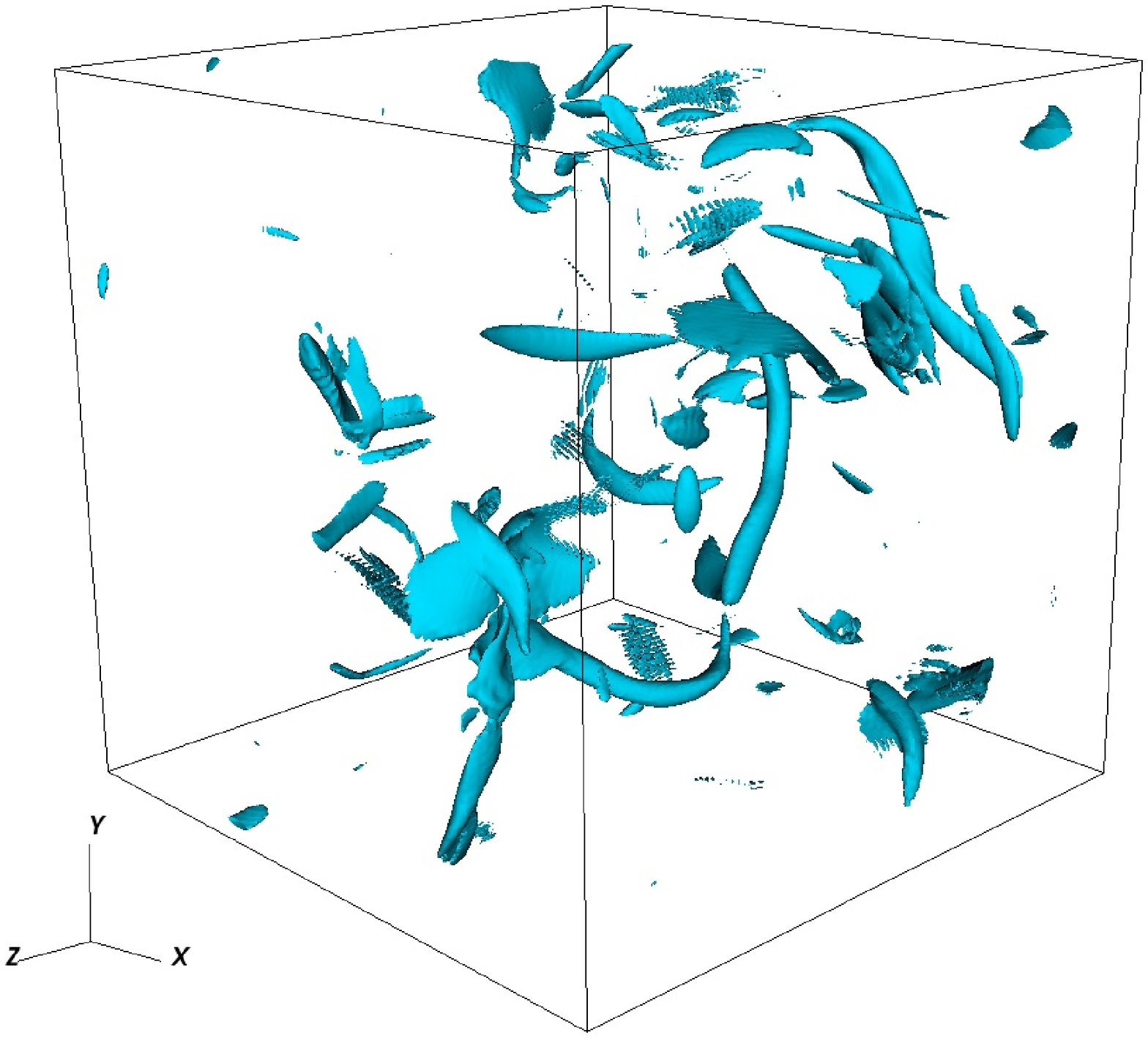}
\end{center}
\caption{\label{figch3:vorforced}{(Color online) Constant-$|\bomega|$ isosurfaces for 
$|\bomega|=\overline{|\bomega|}+2\sigma_\omega$ at $t\approx60\Teddy$ 
without (top panel) and with polymers [middle panel $We=3.5$ ({\tt NSP-256A}) 
and bottom panel $We=7.1$ (run {\tt NSP-256B})];  
$\overline{|\bomega|}$ is the mean  and 
$\sigma_\omega$ the standard deviation of $|\bomega|$.}} 
\end{figure}

\begin{figure}[!h]
\begin{center}
\includegraphics[width=\linewidth]{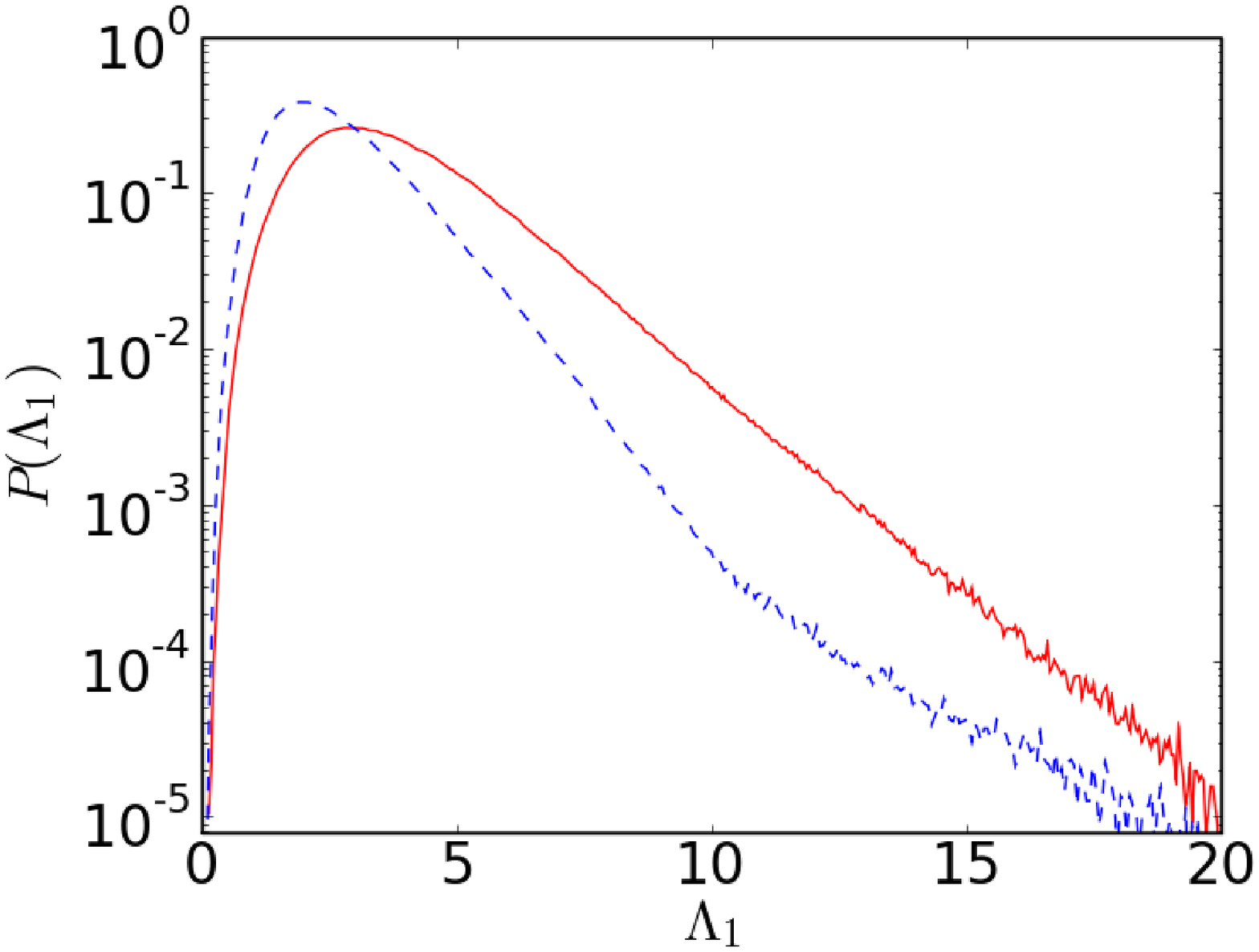}
\end{center}
\caption{\label{figch3:eigsta}{(Color online) Semilog (base 10) plots of the PDF 
$P(\Lambda_1)$ 
versus the first eigenvalue $\Lambda_1$ of the strain-rate tensor $S$  
for the run {$\tt NSP-256B$}, with [$We=7.1$~(blue dashed line)] 
and without [$c=0$~(full red line)] polymer additives. These plots are
normalized such that the area under each curve is unity.}}
\end{figure}

\begin{figure}[!h]
\begin{center}
\includegraphics[width=\linewidth]{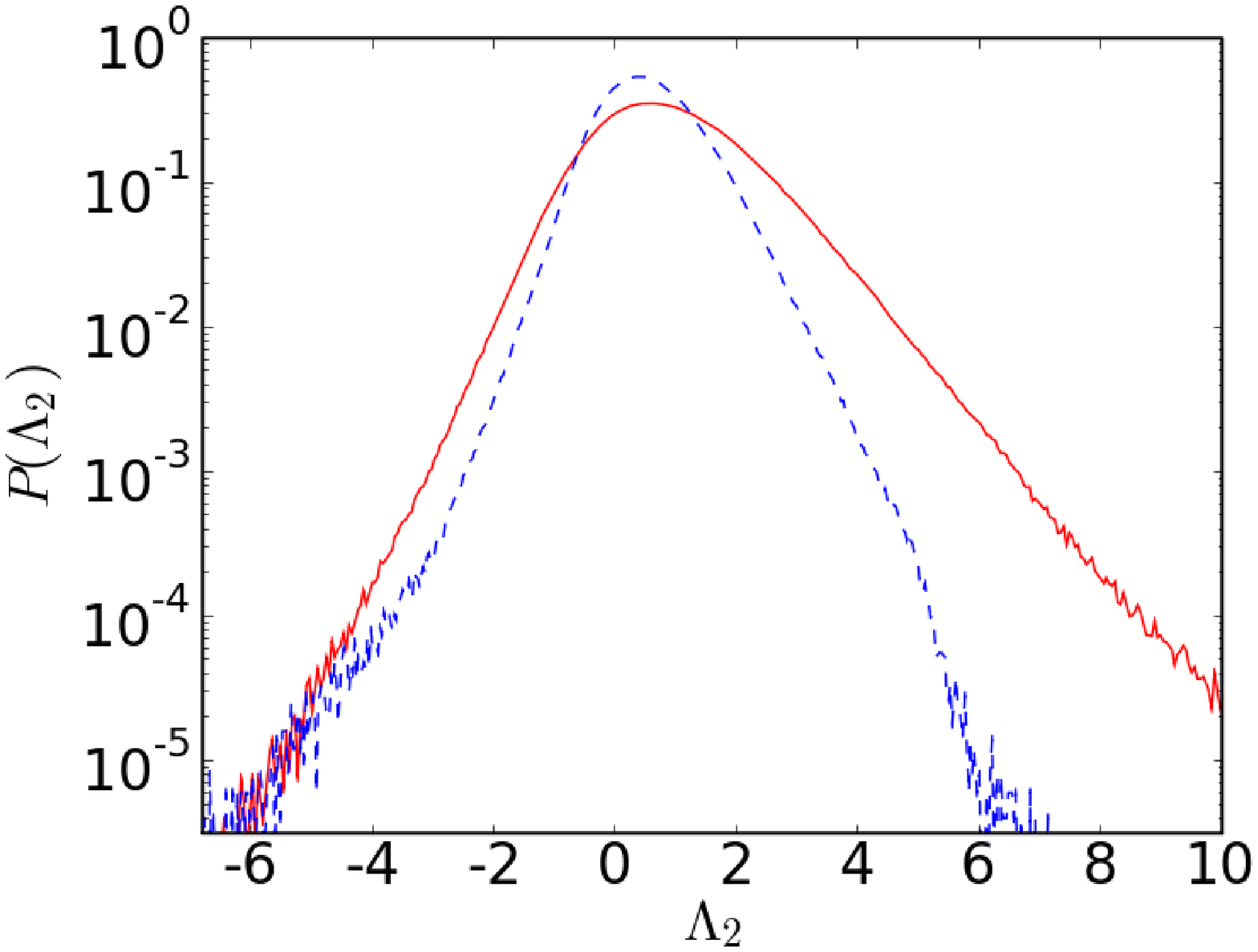}
\end{center}
\caption{\label{figch3:eigstb} {(Color online) Semilog (base 10) plots of the PDF 
$P(\Lambda_2)$ versus the second eigenvalue $\Lambda_2$ of the 
strain-rate tensor $S$  for the run {$\tt NSP-256B$}, with 
[$We=7.1$~(blue dashed line)] and without [$c=0$~(full red line)] 
polymer additives. These plots are normalised such that the area under 
each curve is unity.}}
\end{figure}

\begin{figure*}[]
\begin{center}
\includegraphics[width=0.45\linewidth]{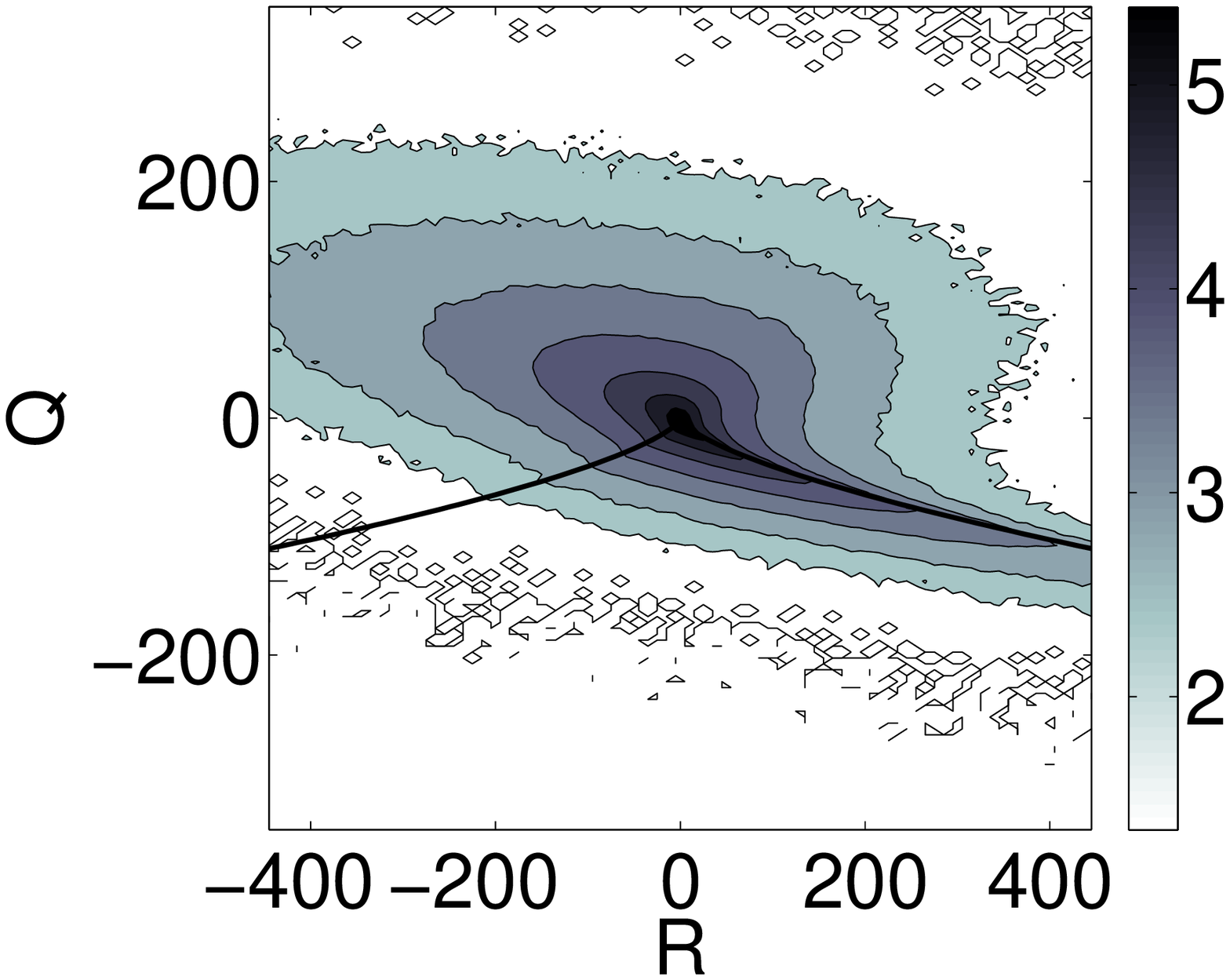}
\includegraphics[width=0.45\linewidth]{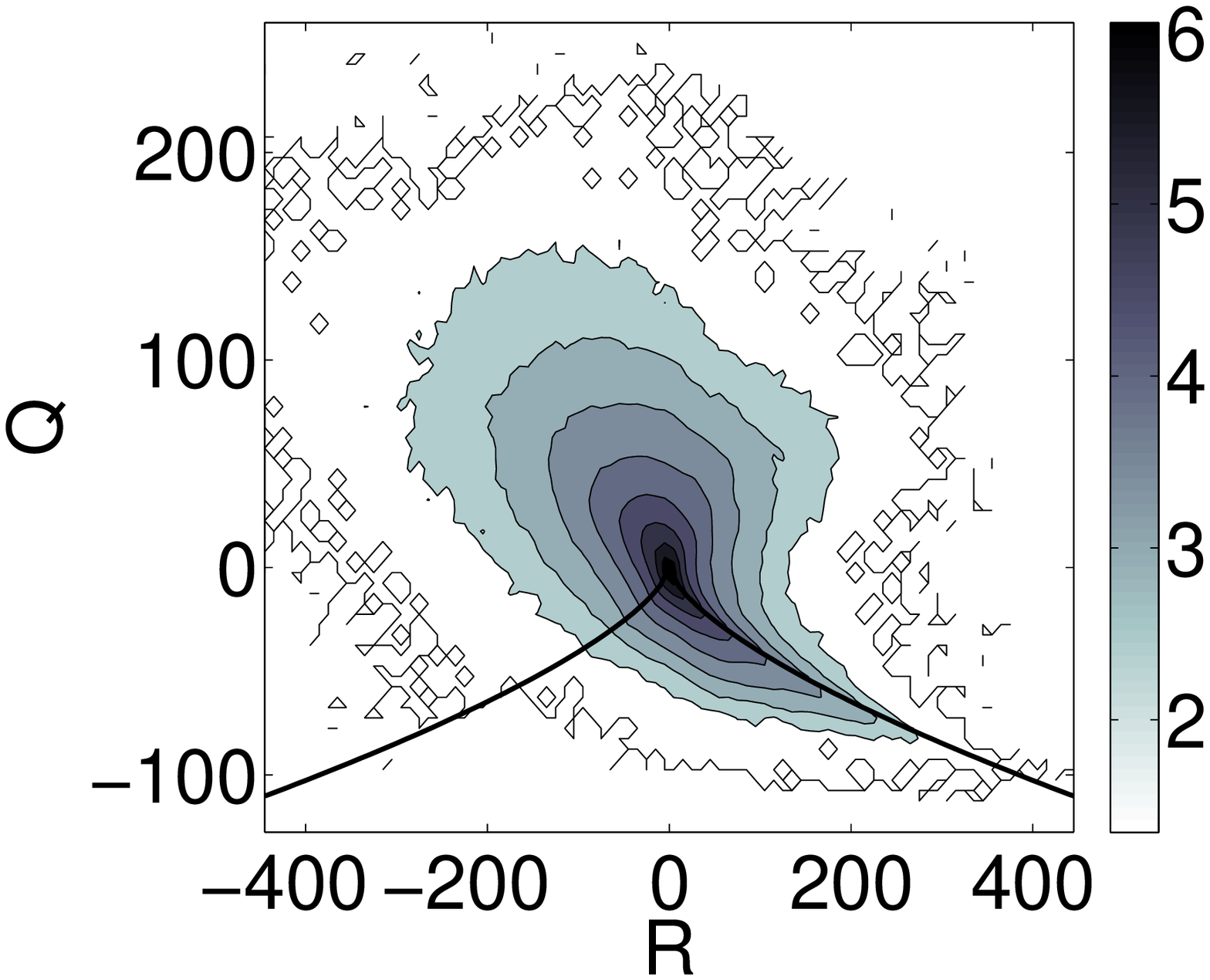}
\end{center}
\caption{\label{figch3:qrp} (Color online) Contour plots of the joint PDF $P(R,Q)$ 
from our DNS studies with (left) and  without (right) polymer additives. 
In this $Q-R$ plot, $Q=-{\rm Tr}(A^2)/2$ and $R=-{\rm
Tr}(A^3)/3$ are the invariants of the velocity-gradient tensor ${\nabla{\bf
u}}$. Note that $P(R,Q)$ shrinks on the addition of polymers; this 
indicates a depletion of small-scale structures. The contour levels are 
logarithmically spaced and are drawn at the following values: 
$1.3, 2.02, 2.69, 3.36, 4.04, 4.70, 5.38$, and $6.05$.}
\end{figure*}

We now investigate how polymers affect small-scale structures in homogeneous,
isotropic, fluid turbulence; and we make specific comparisons with 
experiments~\cite{lib05,lib06}. 
We begin by plotting the PDFs of the modulus of the vorticity $|\bomega|$ and
the local energy dissipation rate $\epsilon_{\rm loc} = \sum_{i,j}
(\partial_i u_j + \partial_j u_i)^2/2$  in Figs.~\ref{figch3:pdfve}. We find
that the addition of polymers reduces regions of high vorticity and high
dissipation [Figs.~(\ref{figch3:pdfve})].  Furthermore, we find that, on
normalising $|\bomega|$ or $\epsilon_{\rm loc}$ by their respective standard
deviations, the PDFs of these normalised quantities for the fluid with and
without polymers collapse onto each other (within our numerical error bars)
as shown in Figs.~\ref{figch3:pdfnorm}.  Our results for these PDFs are in
qualitative agreement with the results of Refs.~\cite{lib05,lib06}~(see
Fig.~$2$ of Ref.~\cite{lib05} and Fig.~$3$ of Ref.~\cite{lib06}).
Earlier high-resolution,
large-$Re_{\lambda}$, DNS studies of homogeneous, isotropic
fluid turbulence {\it without polymer additives}
(see, e.g., Ref.~\cite{kan03,per09b} and references therein) 
have established that iso-$|\bomega|$ surfaces are filamentary for 
large values of $|\bomega|$. In Fig.~\ref{figch3:vorforced} 
 we show how such iso-$|\bomega|$ surfaces change on the addition of 
polymers ($c=0.1$; $We=3.5$ or $7.1$).
In particular, the addition of polymers suppresses a significant
fraction of these filaments (compare the top and middle panels of 
Fig.~\ref{figch3:vorforced}); and this suppression becomes stronger
as $We$ increases (middle and bottom panels of  
Fig.~\ref{figch3:vorforced}).
In addition to suppressing events which contribute to large
fluctuations in the vorticity, the addition of polymers also affects
the statistics of the 
eigenvalues of the rate-of-strain matrix
($S_{ij}=(\partial_i u_j + \partial_j u_i)/\sqrt{2}$), namely,
$\Lambda_n$, with $n=1,2,3$. 
They provide a measure of the local stretching and compression of the fluid. 
In our study, these eigenvalues are arranged 
in decreasing order, i.e.,
$ \Lambda_1 > \Lambda_2 > \Lambda_3$. 
Incompressibility implies that
$\sum_i \Lambda_i = 0$; therefore, 
for an incompressible fluid, one of the eigenvalues ($\Lambda_1$) must be 
positive and one~($\Lambda_3$) negative.  The intermediate 
eigenvalue $\Lambda_2$ can either be positive or negative. In 
Figs.~(\ref{figch3:eigsta}) and (\ref{figch3:eigstb}) we plot the PDFs 
of these eigenvalues. The tails of these PDFs shrink on the addition 
of polymers. This indicates that the addition of the polymers leads to 
a substantial decrease in the regions where there is large strain,  
a result that is in qualitative agreement with the experiments of 
Ref.~\cite{lib05} (see Fig.~$3$(b) of Ref.~\cite{lib05}). 
Evidence for the suppression of small-scale structures 
on the addition of polymers can also be obtained by examining 
the attendant change in the topological properties of a three-dimensional, 
turbulent flow. For incompressible, ideal fluids in three dimensions 
there are two topological invariants: $Q\equiv-{\rm Tr}(A^2)/2$ and 
$R\equiv-{\rm Tr}(A^3)/3$, where $A$ is the velocity-gradient 
tensor ${\nabla{\bf u}}$ 
\footnote{Strictly speaking $Q$ and $R$ are not topological
invariants of the (unforced, inviscid) NSP equations but only of the 
(unforced, inviscid) NS equation.}.
Topological properties of such a flow can be
classified~\cite{per+cho87,can92} by a $Q-R$ plot, which is a 
contour plot of the joint 
probability distribution function (PDF) of $Q$ and $R$.
In Fig.~\ref{figch3:qrp} we give $Q-R$ plots from our DNS studies
with and without polymers; although the qualitative shape of these
joint PDFs remains the same, the regions of large $R$ and $Q$ are 
dramatically reduced on the additions of polymers; this is yet another 
indicator of the suppression of small-scale structures. 
%
\subsection{Effects of polymer additives on deep-dissipation-range spectra}
\label{sub:ekN512} 
\begin{figure}[!h]
\begin{center}
\includegraphics[width=\linewidth]{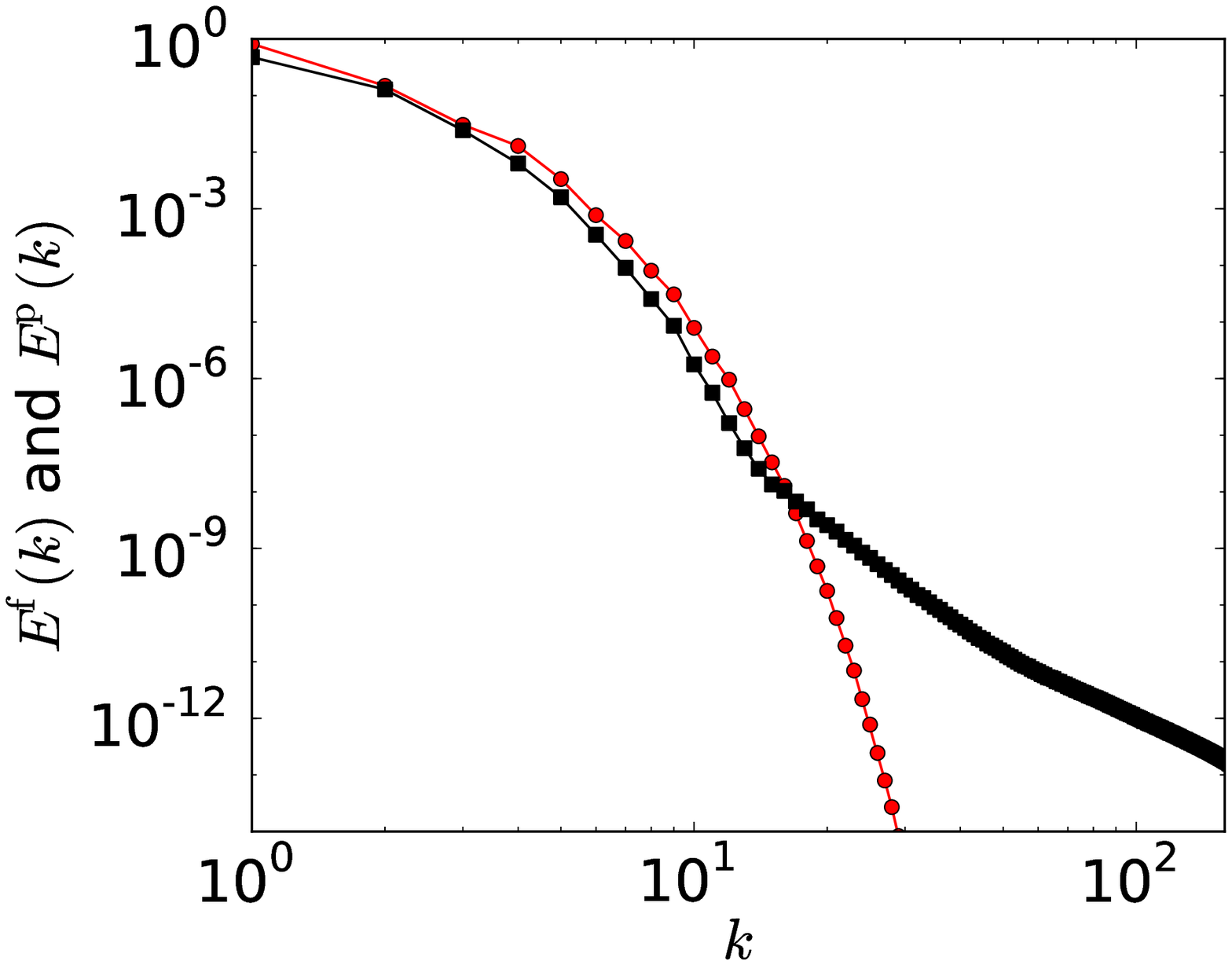}
\end{center}
\caption{\label{figch3:ek} {(Color online) Log-log (base 10) plots of the
fluid-energy spectrum $E^{\rm p}(k)$ versus the magnitude of the wave vector $k$ 
for our run ${\tt NSP=512}$ 
(full black line with squares) for  $c=0.1$ and $\taup=1$. The corresponding plot 
for the pure fluid~(full red line with circles) is also shown for comparison.}}
\end{figure}

\begin{figure}[!h]
\begin{center}
\includegraphics[width=\linewidth]{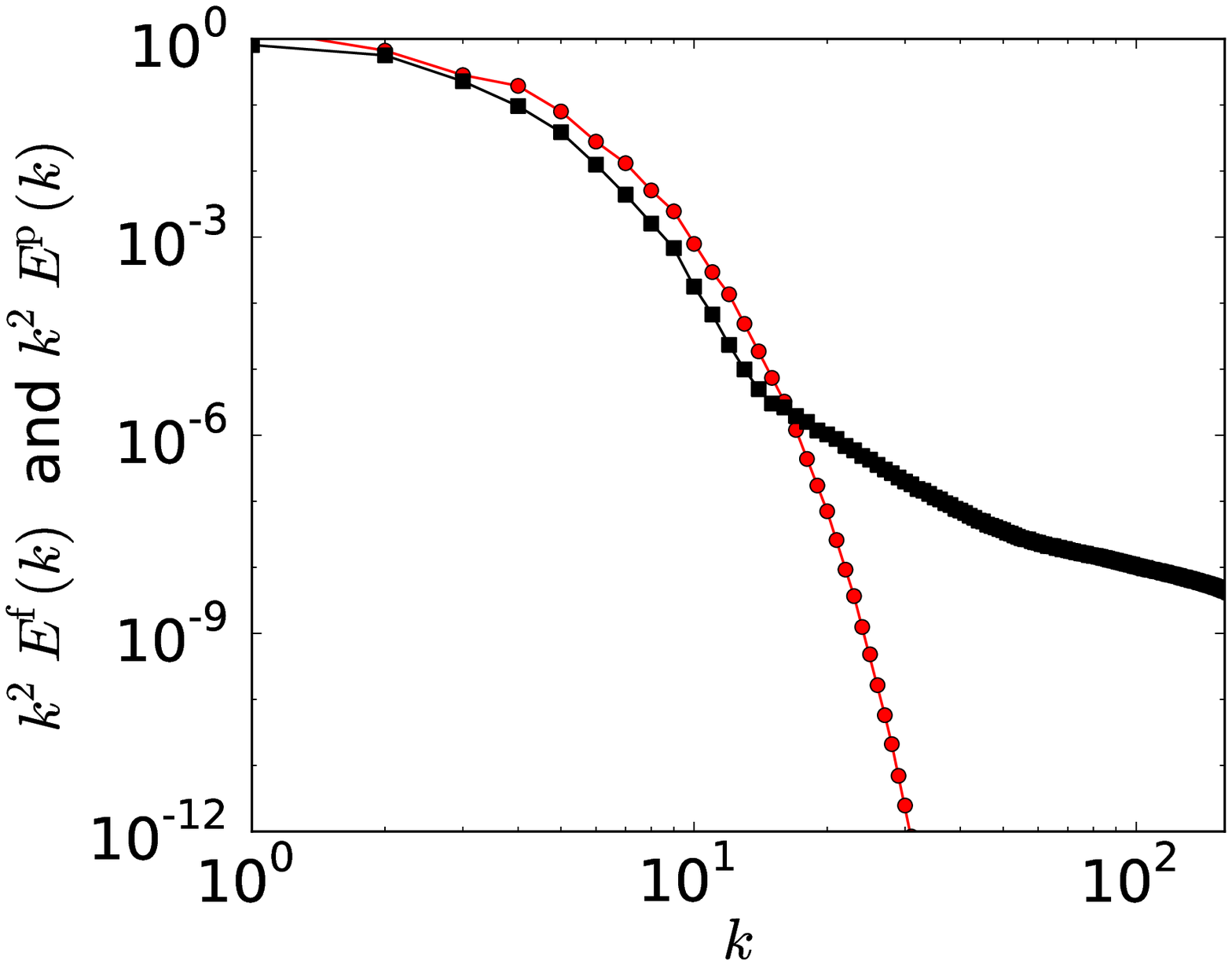}
\end{center}
\caption{\label{figch3:ed} {(Color online) Log-log (base 10) plots of the 
enstrophy spectrum $k^2E^{\rm p}(k)$ versus the magnitude of the wave vector $k$ 
for our run ${\tt NSP=512}$ (full black line with squares) for  $c=0.1$ and 
$\taup=1$. The corresponding plot for the pure fluid~(full red line with circles) is 
also shown for comparison.}}
\end{figure}

In the previous subsections we have studied the effects of polymer additives
on the structural properties of a turbulent fluid at moderate Reynolds 
numbers. We now investigate the effects of polymer additives on the 
deep-dissipation range. To uncover such deep-dissipation-range effects, we 
conduct a very high-resolution, but low-$Re_{\lambda}$ ($=16$) DNS study 
(${\tt NSP-512}$).  The
parameters used in our run ${\tt NSP-512}$ are given in
Table~\ref{tablech3:para}. The fluid is driven by 
using the stochastic-forcing scheme of Ref.~\cite{esw88}.
In Fig.~\ref{figch3:ek} we
plot fluid-energy spectra with and without polymer additives.  The general
behavior of these energy spectra is similar to that in our
decaying-turbulence study~\cite{per06}. We find that, on the addition of
polymers, the energy content at intermediate wave-vectors decreases, whereas
the energy content at large wave-vectors increases significantly. We have
checked explicitly that this increase in the energy spectrum in the
deep-dissipation range is not 
an artifact of aliasing errors: Note first
that this increase starts at wave vectors whose magnitude is considerably
lower than the dealiasing cutoff $k_{\rm max}$ in our DNS; furthermore, the
enstrophy spectrum $k^2E^{\rm p}(k)$, which we plot versus $k$ in
Fig.~\ref{figch3:ed}, decays at large $k$; this indicates that the
dissipation range has been resolved adequately in our DNS.

\begin{figure}[!h]
\begin{center}
\includegraphics[width=\linewidth]{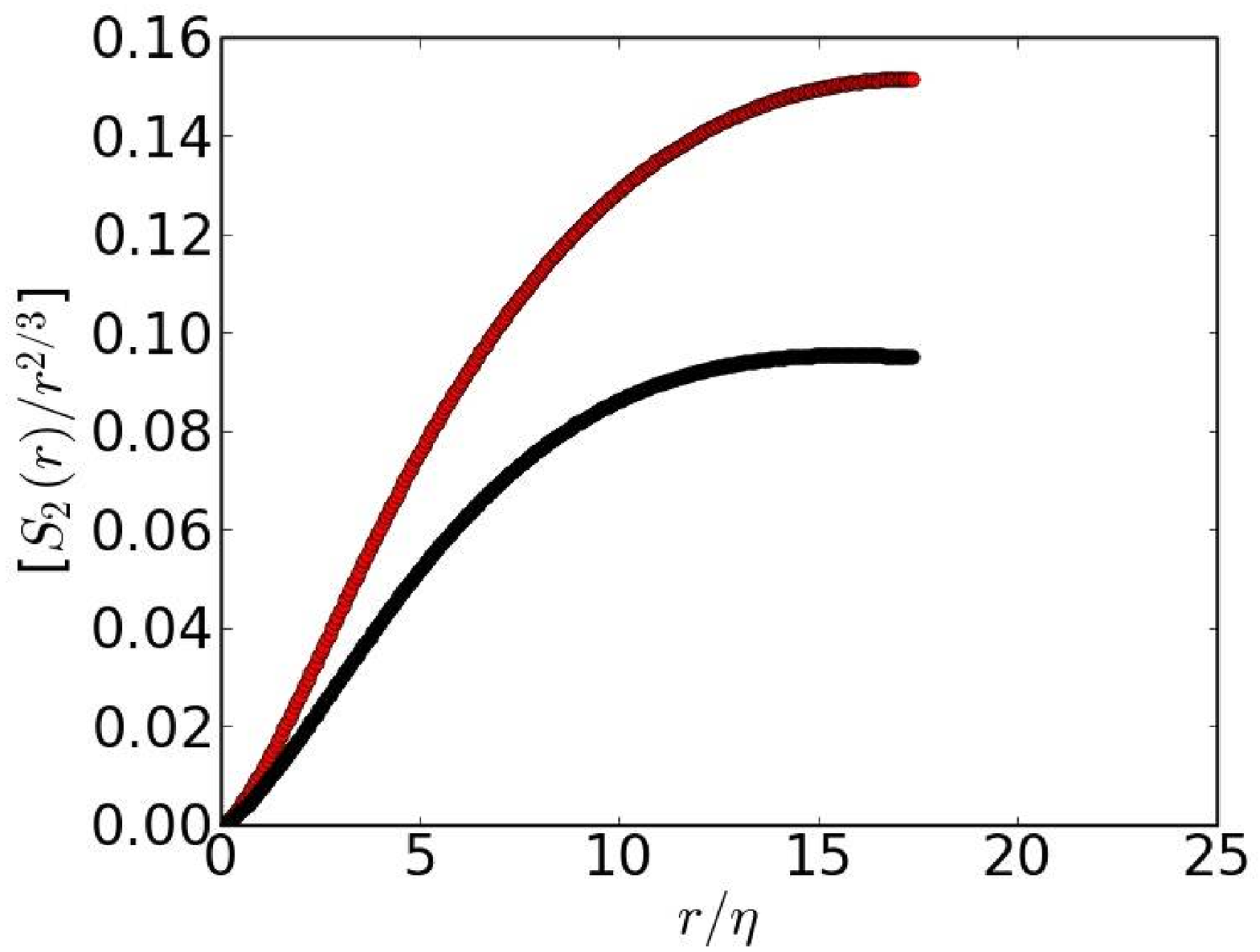}
\end{center}
\caption{\label{figch3:specps} (Color online) Plots of the compensated second-order
structure function $S_2(r)/r^{2/3}$ versus $r/\eta$ with (black circles) 
and without (red circles) polymer additives for our run ${\tt NSP-512}$. 
}
\end{figure}

For homogeneous, isotropic turbulence, the relationship between the 
second-order structure function and the energy spectrum 
is given in Eq.~\eqref{eqch3:s2r}~\cite{Bat53}.
Using this relationship and the data for the energy spectrum shown in 
Fig.~\ref{figch3:ek}, we have
obtained the second-order structure function $S_2(r)$ for our run ${\tt
NSP-512}$. We find that the addition of polymers leads to a decrease in the
magnitude of $S_2(r)$.
Our plots for $S_2(r)$ are similar to those found in the 
experiments of Ref.~\cite{oue09}.
In our simulations we are able to reach much smaller
values of $r/\eta$ than has been possible in experimental studies on these
systems~\cite{oue09}; however, we have not resolved the inertial 
range very well in these runs.
Note that the spectra  $E^{\rm p}(k)$, with polymers, and    
$E^{\rm f}(k)$, without polymers, cross each other as shown 
in Fig.~\ref{figch3:ek}. But such a crossing is not observed in the
corresponding plots  of second-order structure 
functions~(Fig.~\ref{figch3:specps}). 
This can be understood by noting that $S_2(r)$ 
combines large- {\it and} small-$k$ parts~\cite{Dav04} of the
energy spectrum. 
\section{Conclusions}
\label{conclusions}
We have presented an extensive numerical study of the effects
of polymer additives on statistically steady, homogeneous, isotropic 
fluid turbulence. Our study complements, and extends
considerably, our earlier work~\cite{per06}. Furthermore, 
our results compare favorably with several recent experiments. 

Our first set of results show that the average viscous energy dissipation
rate decreases on the addition of polymers. This allows us to extend the
definition of dissipation reduction, introduced in Ref.~\cite{per06}, to the
case of statistically steady, homogeneous, isotropic, fluid turbulence with
polymers. We find that this dissipation reduction increases with an increase
in the Weissenberg number $We$, at fixed polymer concentration $c$. We obtain
PDFs 
of the modulus of the vorticity, of the  eigenvalues $\Lambda_n$ of the
rate-of-strain tensor $S$, and $Q-R$ plots; we find that these
are in qualitative agreement with the experiments of
Refs.~\cite{lib05,lib06}. 

Our second set of results deal with a high-resolution DNS that we have
carried out to elucidate the deep-dissipation-range forms of (a) energy
spectra and (b) the related second-order velocity structure functions. 
We find that this deep-dissipation-range behavior is akin to that in  
our earlier DNS of decaying, homogeneous, isotropic, fluid turbulence
with polymers~\cite{per06}. Furthermore, the results we obtain for the
scaled, second-order, velocity structure $S_2(r)$ yield trends
that are in qualitative agreement with the experiments of Ref.~\cite{oue09}.

We hope that the comprehensive study that we have presented here will
stimulate further detailed experimental studies of the statistical properties
of homogeneous, isotropic fluid turbulence with polymer additives.
\section{Acknowledgments}
\label{acknowledgments}
We thank J. Bec, E. Bodenschatz, F. Toschi, and H. Xu for discussions,
Leverhume trust, 
European Research Council under the AstroDyn Research Project No.\ 227952,
CSIR, DST, and UGC (India) for support, and SERC (IISc) for 
computational resources. Two of us (RP and PP) are members of
the International Collaboration for Turbulence Research and they 
acknowledge support from the COST Action MP0806. 
While most of this work is being carried out PP was a student at IISc. 
%
%


\begin{thebibliography}{10}
\bibitem{hoy77}
J. Hoyt and J. Taylor, Phys. Fluids. {\bf 20},  S253  (1977).

\bibitem{kal_poly04}
C. Kalelkar, R. Govindarajan, and R. Pandit, Phys.~Rev. E {\bf 72},  017301
  (2004).

\bibitem{per06}
P. Perlekar, D. Mitra, and R. Pandit, Phys. Rev. Lett. {\bf 97},  264501
  (2006).

\bibitem{toms49}
B. Toms,  in {\em Proceedings of First International Congress on Rheology}
  (North-Holland, Amsterdam, 1949), pp.\ Section II, 135.

\bibitem{lum73}
J. Lumley, J. Polymer Sci {\bf 7},  263  (1973).

\bibitem{vir75}
P. Virk, AIChE {\bf 21},  625  (1975).

\bibitem{too97}
J.M.J.~Den~Toonder, M.A.~Hulsen, G.D.C.~Kuiken and F. Nieuwstadt, 
J. Fluid. Mech. {\bf 337},  193  (1997).

\bibitem{dub01}
Y. Dubief and S. Lele, Annual research briefs, Center for Turbulence Research
  (unpublished).

\bibitem{pta01}
P. Ptasinski, F. Nieuwstadt, B.~V. den Brule, and M. Hulsen, Flows, Turbulence
  and Combustion {\bf 66},  159  (2001).

\bibitem{pta03}
P. Ptasinski {\it et~al.}, J.~Fluid.~Mech {\bf 490},  251  (2003).

\bibitem{lvo04}
V. L'vov, A. Pomyalov, I. Procaccia, and V. Tiberkevich, Phys. Rev. Lett. {\bf
  92},  244503  (2004).

\bibitem{ang05}
E. {De Angelis}, C. Casicola, R. Benzi, and R. Piva, J.~Fluid.~Mech {\bf 531},
  1  (2005).

\bibitem{pro08}
I. Procaccia, V. \'{L}vov, and R. Benzi, Rev.~Mod.~Phys. {\bf 80},  225
  (2008).

\bibitem{oue09}
N. Ouellette, H. Xu, and E. Bodenschatz, J. Fluid Mech. {\bf 629},  375
  (2009).
A.M~Crawford, N.~Mordant, H. Xu and E.~Bodenschatz,
New Journal of Physics {\bf 10} (2008) 123015.
A.M~Crawford, N.~Mordant, A.La Porta and E.~Bodenschatz,
in {\em Advances in Turbulence IX, Ninth European Turbulence
Conference}, I.P. Castro, P.E. Hancock and T.G. Thomas (Eds),
CIMNE, Barcelona, 2002.

\bibitem{lib05}
A. Liberzon {\it et~al.}, Phys. Fluids. {\bf 17},  031707  (2005).

\bibitem{lib06}
A. Liberzon, M. Guala, W. Kinzelbach, and A. Tsinober, Phys. Fluids. {\bf 18},
  125101  (2006).

\bibitem{ben03}
R. Benzi, E. {De Angelis}, R. Govindarajan, and I. Procaccia, Phys.~Rev.~E {\bf
  68},  016308  (2003).

\bibitem{vai03}
T. Vaithianathan and L. Collins, Journal of Computational Physics {\bf 187},  1
   (2003).

\bibitem{lam05}
A. Lamorgese, D. Caughey, and S. Pope, Phys. Fluids {\bf 17},  015106  (2005).

\bibitem{esw88}
V. Eswaran and S. Pope, Computers and Fluids {\bf 16},  257  (1988).

\bibitem{Can88}
C. Canuto, M. Hussaini, A. Quarteroni, and T. Zang, {\em Spectral methods in
  Fluid Dynamics} (Spinger-Verlag, Berlin, 1988).

\bibitem{vin91}
A. Vincent and M. Meneguzzi, J.~Fluid.~Mech. {\bf 225},  1  (1991).

\bibitem{vai06}
T. Vaithianathann, A. Robert, J. Brasseur, and L. Collins, J. Non-newtonian
  Fluid Mech. {\bf 140},  3  (2006).

\bibitem{kt00}
A. Kurganov and E. Tadmor, J. Comp. Phys. {\bf 160},  241  (2000).

\bibitem{Bat53}
G. Batchelor, {\em The theory of homogeneous turbulence} (Cambridge University
  Press, Cambridge, 1953).

\bibitem{sch+sre+yak89}
J. Schumacher, K.R. Sreenivasan and V. Yakhot, 
New J. Phys., {\bf 9}, 89 (2007). 

\bibitem{rob+vai+col+bra10}
A. Robert, T. Vaithianathan, L. Collins, and J. Brasseur, Journal of Fluid
  Mechanics,  in press, (2010).

\bibitem{mit05a}
D. Mitra, J. Bec, R. Pandit, and U. Frisch, Phys.~Rev.~Lett {\bf 94},  194501
  (2005).

\bibitem{kan03}
Y. Kaneda {\it et~al.}, Physics of Fluids {\bf 15},  L21  (2003).

\bibitem{per09b}
R. Pandit, P. Perlekar, and S.S. Ray, Pramana {\bf 73},  179  (2009).

\bibitem{per+cho87}
A. Perry and M. Chong, Annu. Rev. Flu. Mech. {\bf 19},  125  (1987).

\bibitem{can92}
B. Cantwell, Phys. Fluids A {\bf 4},  782  (1992).


\bibitem{Dav04}
P. Davidson, {\em Turbulence} (Oxford University Press, New York, 2004), pp.\
  386--410.

\end{thebibliography}
\end{document}